\documentclass[nofootinbib, reprint,superscriptaddress,amsmath,amssymb,floatfix,pra,hidelinks]{revtex4-2}

\usepackage{orcidlink}
\usepackage{xcolor}
\usepackage{graphicx}
\usepackage{braket}
\usepackage{mathtools}
\usepackage{tikz}
\usepackage{quantikz}
\usepackage[caption=false,font=small]{subfig} 
\usepackage{pgffor}
\usepackage{tabularx}
\usepackage[a4paper,margin=2.2cm]{geometry}
\usepackage{amsmath,amssymb,physics,braket}
\usepackage{graphicx}
\usepackage{float}
\usepackage{dsfont}
\usepackage{svg}
\usepackage{pgfplots}
\usepackage{adjustbox} 
\usepackage{xcolor} 
\usepackage{cleveref}
\crefname{figure}{Fig.}{Figs.}
\crefname{equation}{Eq.}{Eqs.}

\begin{document}

\title{Ultimate sensitivity of multiparameter estimation in quantum sensing with undetected photons}
\author{Sanjeet Swaroop Panda\,\orcidlink{0000-0002-3419-7150}}
\email{sanjeet.panda@anu.edu.au}
\affiliation{ Department of Quantum Science and Technology, Australian National University, Canberra, ACT 2601, Australia}
\affiliation{A*STAR Quantum Innovation Centre (Q.InC), Agency for Science, Technology and Research (A*STAR), 2 Fusionopolis Way, 08-03 Innovis, 138634, Singapore}
\author{Lorc\'an O. Conlon\,\orcidlink{0000-0002-0921-5003}}
\affiliation{A*STAR Quantum Innovation Centre (Q.InC), Agency for Science, Technology and Research (A*STAR), 2 Fusionopolis Way, 08-03 Innovis, 138634, Singapore}
\affiliation{Joint Quantum Institute and Joint Center for Quantum Information and Computer Science, NIST/University of Maryland, College Park, Maryland 20742, USA}
\affiliation{Centre for Quantum Technologies, National University of Singapore, 3 Science Drive 2, 117543, Singapore}
\author{Li Gong\,\orcidlink{0009-0008-8499-4562}}
\affiliation{A*STAR Quantum Innovation Centre (Q.InC), Agency for Science, Technology and Research (A*STAR), 2 Fusionopolis Way, 08-03 Innovis, 138634, Singapore}
\author{Ping Koy Lam\,\orcidlink{0000-0002-4421-601X}}
\affiliation{A*STAR Quantum Innovation Centre (Q.InC), Agency for Science, Technology and Research (A*STAR), 2 Fusionopolis Way, 08-03 Innovis, 138634, Singapore}
\affiliation{Centre for Quantum Technologies, National University of Singapore, 3 Science Drive 2, 117543, Singapore}
\author{Jie Zhao\,\orcidlink{0000-0002-7382-1964}}
\affiliation{ Department of Quantum Science and Technology, Australian National University, Canberra, ACT 2601, Australia}
\author{Young-Wook Cho\,\orcidlink{0000-0002-9673-3916}}
\affiliation{A*STAR Quantum Innovation Centre (Q.InC), Agency for Science, Technology and Research (A*STAR), 2 Fusionopolis Way, 08-03 Innovis, 138634, Singapore}
\affiliation{Centre for Quantum Technologies, National University of Singapore, 3 Science Drive 2, 117543, Singapore}
\author{Ruvi Lecamwasam\,\orcidlink{0000-0001-6531-3233}}
\email{me@ruvi.blog}
\affiliation{A*STAR Quantum Innovation Centre (Q.InC), Agency for Science, Technology and Research (A*STAR), 2 Fusionopolis Way, 08-03 Innovis, 138634, Singapore}
\affiliation{Centre for Quantum Technologies, National University of Singapore, 3 Science Drive 2, 117543, Singapore}
\date{\today}

\begin{abstract}
Quantum sensing with undetected photons is a technique where photons of one wavelength probe a sample, but information is extracted by measuring photons of another wavelength that never interacts with the sample. This has seen significant experimental advances in applications such as spectroscopy, microscopy, and bio-sensing. However, a detailed theoretical analysis using the tools of quantum metrology is currently lacking. Thus it is unclear how far away current schemes are from fundamental limits, and what the optimal measurement strategies are. We apply a multiparameter quantum estimation framework to quantify the error when estimating the unknown transmission and phase shift of a sample. The optimal measurement scheme is shown to require only a single controllable phase shift, easily implementable in existing setups. 
We also study how to use multipass interactions to maximise information gain. In general the optimum number of passes scales inversely with the log of the transmission of the sample. 
This work clarifies the metrological power of quantum sensing with undetected photons, and provides guidance for the design of experiments requiring high sensitivity.
\end{abstract}

\maketitle

\section{Introduction}
\label{sec:intro}

Quantum sensors use quantum phenomena to gain information beyond the capability of classical sensors \cite{degen2017quantum,Giovannetti2011,Demkowicz2015}. 
Most previous works in quantum sensing focuses on the potential of phenomena such as entanglement \cite{giovannetti2006quantum,nagata2007beating} and squeezing \cite{lawrie2019quantum,mcculler2020frequency} to exhibit enhanced precision. However, the fundamental nature of quantum systems can provide advantages in other ways. While atomic magnetometers and gravimeters can exhibit state of the art sensitivity \cite{wang2025pulsed,dickerson2013}, they are finding commercial application due to the inherently low drift of atomic systems, compared to mechanical sensors whose properties vary with temperature and wear \cite{fang2024classical,bennett2021precision,muradoglu2025quantum}. Another area that has seen extensive development is quantum sensing with undetected photons \cite{lemos2014}. In this technique a nonlinear interferometer is formed using photons of two wavelengths. One wavelength passes through a sample, but information about the sample is extracted by measuring the other wavelength \cite{ZouMandel1991,WisemanMolmer2000,RevModPhys.94.025007}. This provides tremendous flexibility, by allowing for separate engineering of the optical fields used for probing and detection.

A key thrust of experimental efforts has been in imaging and spectroscopy in the infrared regime. At these wavelengths, coherent sources and efficient detectors can be prohibitively expensive, or require cryogenic cooling for high efficiency. Imaging with undetected photons allows a sample to be probed with infrared light generated by spontaneous parametric detection, while the signal is extracted using high-efficiency visible light detectors.  This has been used for spectroscopy and chemical identification \cite{Kalashnikov2016,paterova2017nonlinear}, optical coherence tomography \cite{paterova2018tunable}, microscopy \cite{Kviatkovsky2020,suryana2025infrared}, and gas sensing \cite{sabanin2025sensing}.
More recent work has focused on the manipulation and characterization of the underlying quantum states themselves, including experimental distillation of quantum imaging \cite{fuenzalida2023experimental} and full quantum state tomography of the undetected idler photons \cite{Fuenzalida_2024, sciadv}.

Despite this wealth of experimental progress, there has been very little study of this problem using the tools of quantum metrology. Thus fundamental limits, and optimal measurement schemes, are unknown. 
Only recently has sensing with undetected photons been analysed using quantum Fisher information (QFI) \cite{zhang2025measurementuncertaintyinfraredspectroscopy,houde2026quantum}. 
The QFI is an important tool in single parameter estimation problems, where it can find fundamental limits, and design optimal measurement schemes \cite{BraunsteinCaves1994,Helstrom1976,Paris2009,Liu_2019}. However, it faces limitations when we must estimate two or more unknown quantities simultaneously. This is the case in quantum sensing with undetected photons, where both the transmission and phase shift of the sample are unknown. In this case, the precision suggested by the QFI is often not achievable \cite{conlon2024roleextendedhilbertspace,conlon2024gappersistencetheoremquantum,Yang_2019,Szczykulska2016, imai2026hierarchysaturationconditionsmultiparameter}. Instead we must use more sophisticated tools such as the Holevo \cite{HolevoBook2011} and Nagaoka Cram\'er-Rao bounds \cite{Nagaoka}.

In addition to the complexities of multiparameter estimation, most applications of quantum sensing are for measurement of very weak signals, such as in biology or gas sensing. A common method of amplifying weak signals is multipass interactions, where the probe passes through the sample more than once. 
However increasing the number of passes arbitrarily will eventually degrade the signal, due to unavoidable losses that build up with each interaction with the sample.\footnote{Under certain noise models it is known that the effect of noise can be removed via a error correction scheme\,\cite{Zhou_2018}. However, we do not consider such schemes here.} The optimum number of passes for sensing with undetected photons is currently unknown. 

In this work we analyse quantum sensing with undetected photons from a rigorous multiparameter estimation framework. In \cref{sec:multipass_qici} we introduce a simple physical model, and necessary background on multiparameter estimation theory. In \cref{sec:SingleParameter} this is applied to first consider the case of single-parameter estimation. We derive and analyse the optimal measurement schemes for single parameter estimation, and compare these with the detection scheme used in existing experiments. We also see how this changes with multiple passes. In \cref{sec:MultiparameterEstimation} we then apply the Holevo and Nagaoka bounds to study the multiparameter estimation case. Using these we find the fundamental limits of sensing with undetected photons, and construct optimal measurement schemes to achieve these.

Together, these results clarify the metrological power of quantum sensing with undetected photons. These highlight precisely when and in what manner a multipass geometry can provide genuine advantage, and provide concrete design rules for future experiments requiring high-sensitivity characterization of sample transmission and phase shift.

\section{Background}
\label{sec:multipass_qici}

\subsection{Model for sensing with undetected photons}
\label{sec:qmodel_3state}

\begin{figure*}[t]
  \centering
    \includegraphics[width=\textwidth]{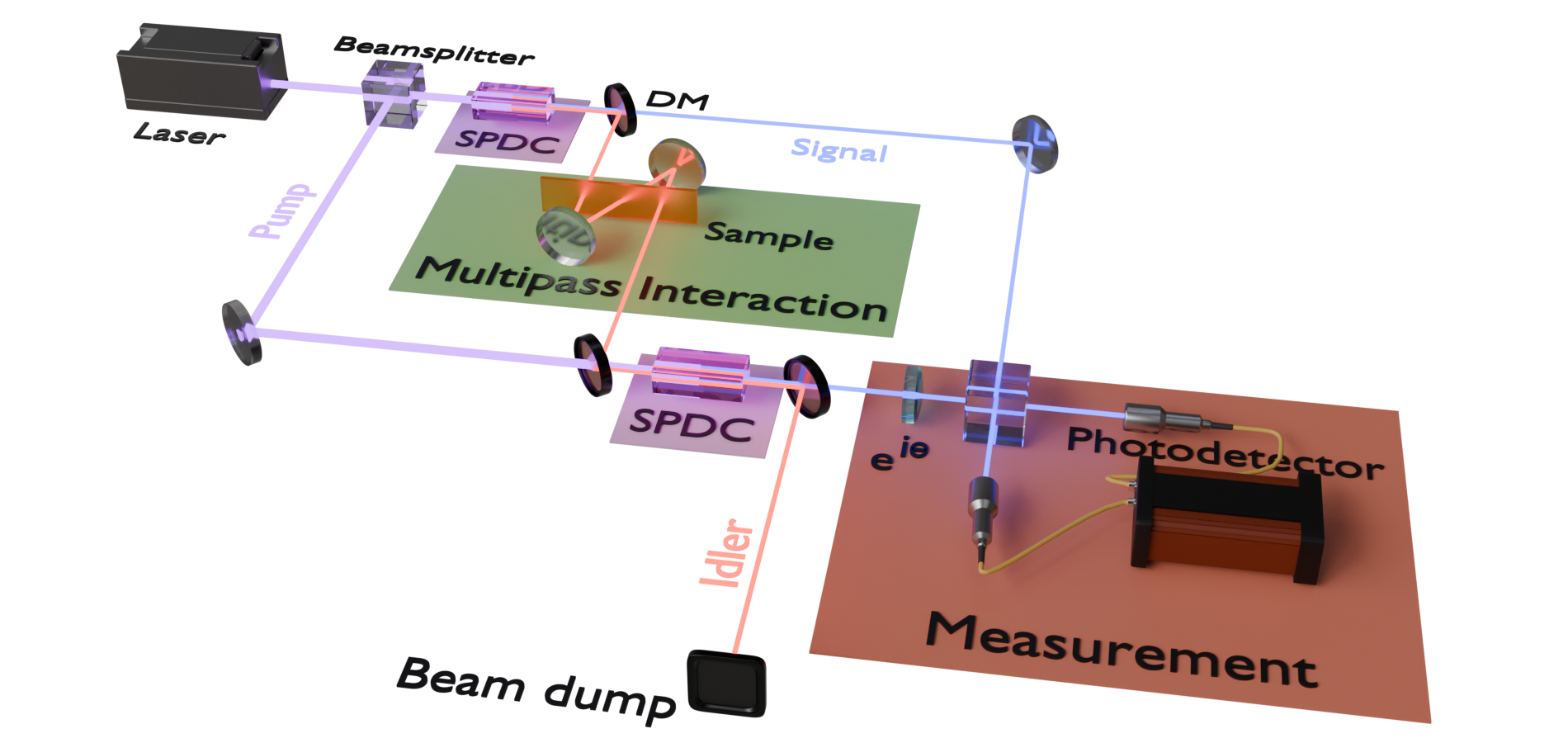}
  \caption{Schematic for quantum sensing with undetected photons \cite{Barreto_Lemos_2022}. A pump laser drives two SPDC sources, which convert laser photons to signal and idler photons. The idler from the first SPDC source is passed through a sample one or more times, and then directed through the second SPDC source. The final idler photon, consisting of infrared photons from either SPDC source, is not measured. It may be released to the environment, or passed into a beam dump as shown in this diagram. The signal photons have a relative phase shift of $e^{i\theta}$ introduced, and then are interfered and measured. Due to induced coherence, these show interference depending on the properties of the sample. In many applications the idler photon is infrared light while the signal photon is visible, allowing for infrared sensing using high-efficiency visible light photodetectors. In the figure `DM' denotes dichroic mirrors, which reflect only the idler photons while transmitting the pump and signal photons.}\label{fig:QIUPSetup}
  
\end{figure*}

In Quantum Sensing with Undetected Photons (QSUP), a sample is probed by photons of one wavelength, while information is extracted by detecting photons of another wavelength that do not directly interact with the sample. These photons are generated using Spontaneous Parametric Down-Conversion (SPDC), where a nonlinear crystal `down-converts' the laser photons into two photons of lower energy. 
The standard layout is shown in \cref{fig:QIUPSetup}, consisting of a nonlinear interferometer formed by two SPDC sources. Following standard terminology, we will refer to the photons which probe the sample as idlers, and the photons that are measured as the signal. In many imaging applications, the idler photon is infrared while the signal photon is visible. After passing through the sample, the idler is mode-matched with the second SPDC source, so that idler photons produced by either source are indistinguishable. This induces coherence between signal photons from the two sources. Thus we can image the sample by interfering the signal photons, even though none of the signal photons directly interacted with the sample. Prior to interference, a controllable phase shift $e^{i\theta}$ is introduced between the signal photons. We note that other configurations of quantum  with undetected photons  exist\, \cite{Barreto_Lemos_2022}, such as folded setups making use of a single SPDC source\,\cite{paterova2017nonlinear}. However, our analysis remains applicable.

We allow for a multi-pass interaction, where the idler photon passes through the sample multiple times. This can increase the sensitivity, by increasing the phase and amplitude shift imparted by the sample. However, there is a trade-off. With each pass the sample absorbs some of the idler photons, reducing visibility of the final interference pattern.

We will now model the system shown in \cref{fig:QIUPSetup}, following \cite{lemos2014,Barreto_Lemos_2022}. We begin by enumerating the different modes in our Hilbert space. Each SPDC source emits a signal and idler mode. However, the two idler modes are made degenerate through optical alignment. We will also consider an environmental mode, where photons get scattered to when they are absorbed by the sample. Thus the basis elements of our Hilbert space are of the form
\begin{equation}
    |a\rangle_{s_1}|b\rangle_{s_2}|c\rangle_{i}\ket{d}_E,
\end{equation}
where $|a\rangle_{s_1}$ is the state of the signal mode from the first crystal, $|b\rangle_{s_2}$ the signal mode from the second crystal, $|c\rangle_i$ the degenerate idler mode driven by both crystals, and $\ket{d}_E$ the environment.

Let us consider the state of the system prior to entering the `Measurement' block in \cref{fig:QIUPSetup}. If a photon pair was generated by the first SPDC crystal, the state is
\begin{equation}
    \begin{aligned}
        \lvert\psi_1\rangle &=|1\rangle_{s_1}|0\rangle_{s_2}\left(t^{n} e^{i n\phi}\,\ket{1}_{i_1}\!\otimes\!\ket{0}_E\vphantom{\sqrt{1-t^{2n}}} \right. \\
        &\left.\hspace{7em}+ \sqrt{1-t^{2n}}\,\ket{0}_{i_1}\!\otimes\!\ket{1}_E \right),
    \end{aligned}
\end{equation}
where $t$ and $\phi$ are the transmission and phase shift induced by the sample, and $n$ is the number of passes. If a photon pair was produced by the second crystal, the state is
\begin{equation}
    \lvert\psi_2\rangle=|0\rangle_{s_1}|1\rangle_{s_2}|1\rangle_i\ket{0}_E.
\end{equation}
Since each SPDC source can generate a photon pair with equal probability, the state will be a superposition of the two:
\begin{equation}
    |\Psi\rangle=\frac{\lvert\psi_1\rangle+\lvert\psi_2\rangle}{\sqrt{2}}.
\end{equation}
Neither the idler nor environmental mode are measured. Thus the state of the signal modes prior to the measurement stage is given by the partial trace over the idler and environmental modes:
\begin{equation}
    \rho=\mathrm{tr}_{i, E}\left\{\lvert\Psi\rangle\langle\Psi\rvert\right\}.
\end{equation}
This density matrix describes the state of the two signal modes.

We provide detailed calculations in \S I of the Supplementary Material. There we see that the system is described by a two mode Hilbert space with basis
\begin{equation}
    \begin{aligned}
        |0\rangle &= |1\rangle_{s_1}|0\rangle_{s_2}, \\
        |1\rangle &= |0\rangle_{s_1}|1\rangle_{s_2}.
    \end{aligned}
\end{equation}
With respect to this basis, the state of the signal modes is
\begin{equation}
    \label{eq:rho_signal_3state}
    \rho =\frac12
        \begin{pmatrix}
        1 & t^{n}e^{-in\phi}\\[3pt]
        t^{n}e^{+in\phi} & 1
        \end{pmatrix}.
\end{equation}

We observe that the sample transmittance $t$ directly determines the purity of the reduced output state, whereas the phase $\phi$ does not. The effect of multiple passes is a rescaling $t \mapsto t^{\,n}$ and $\phi \mapsto n\phi$.

Let us now consider measurement of $\rho$. This consists of a relative phase shift $e^{i\theta}$, followed by a 50:50 beamsplitter, and then photodetection. This is equivalent to a projective measurement of $\rho$ in the  basis
\begin{equation}
    \ket{\pm_\theta}=\frac{\ket{0}\pm e^{i\theta}\ket{1}}{\sqrt{2}}.
\end{equation}
The photon intensity at the photodetectors is then
\begin{equation}\label{eq:ModelDetectionProbabilities}
    \mathrm{tr}\left\{\rho\,|\pm_{\theta}\rangle\langle \pm_{\theta}|\right\}=\frac{1}{2}\left(1\pm t^n\cos(\theta + n\phi)\right).
\end{equation}

Thus as $\theta$ or $\phi$ are varied, we will observe an interference pattern with fringe visibility $t^n$. This interference pattern in the signal photons provides information about the sample, despite the fact that it was only the `undetected' idler photon mode which interacted with the sample.

\subsection{Fundamental limits to sensing}
We wish to find the ultimate limits to sensitivity of QSUP, and the best measurement schemes to achieve this. To this end, we will briefly introduce some tools from quantum information and multiparameter estimation theory.

\subsubsection{Quantum Fisher Information}

In quantum parameter estimation, a parameter $\theta$ is encoded into a quantum state $\rho_\theta$. After measuring the output, one forms an estimator $\hat{\theta}$ from the data. Intuitively, the Quantum Fisher Information (QFI) quantifies how rapidly $\rho_\theta$ changes with $\theta$: if $\rho_{\theta}$ and $\rho_{\theta+d\theta}$ become distinguishable quickly, then small parameter shifts can be estimated more precisely. The QFI itself is thus independent of measurement scheme, and provides a bound on the information gained by any measurement. In the case of single parameter estimation, this bound is always achievable. Thus the quantum Fisher information is widely used to benchmark quantum sensors, optimize probe states and resources, and quantify the impact of noise and loss
 \cite{Demkowicz2015,Giovannetti2011}.
    
Let us first consider estimation of a single unknown parameter $\theta$. For any fixed POVM $\{\Pi_x\}$ with outcome probabilities
\begin{equation}
        p(x|\theta)=\mathrm{Tr}\!\left(\rho_\theta\,\Pi_x\right),
        \label{eq:px_theta}
\end{equation}
the \emph{classical Fisher information} (CFI) is
    \begin{equation}
        F_C(\theta)=\sum_x \frac{\bigl[\partial_\theta p(x|\theta)\bigr]^2}{p(x|\theta)},
        \label{eq:cfi_def}
    \end{equation}
with the sum replaced by an integral if the set of outcomes is continuous.  The classical Cram\'er--Rao bound states that if we measure the probability distribution $m$ times, the variance of any locally unbiased estimator $\hat{\theta}$ for $\theta$ is bounded as
\begin{equation}
    \mathrm{Var}(\hat{\theta}) \;\ge\; \frac{1}{m\,F_C(\theta)}.
                \label{eq:crb_classical}
\end{equation}
The \emph{quantum Fisher information} (QFI), denoted $F_Q(\theta)$, is the maximum Fisher information attainable over \emph{all} possible measurements. 
\begin{equation}
        F_Q(\theta)=\max_{\{\Pi_x\}} F_C(\theta).
\end{equation}
    This leads to the \emph{quantum Cram\'er--Rao bound} (QCRB),
\begin{equation}
        \mathrm{Var}(\hat{\theta}) \;\ge\; \frac{1}{m\,F_C(\theta)} \;\ge\; \frac{1}{m\,F_Q(\theta)},
        \label{eq:qcrb}
\end{equation}
which provides a measurement-independent benchmark for the best possible local precision. 
    
We can compute the quantum Fisher information using the Symmetric Logarithmic Derivative (SLD) $L_{\theta}$, defined implicitly as \cite{Liu_2019}
\begin{equation}\label{eq:SLDDefinition}
    \partial_{\theta}\rho=\frac{1}{2}\left(L_{\theta}\rho+\rho L_{\theta}\right),
\end{equation}
where $\rho$ is the quantum state being measured. In terms of this, the quantum Fisher information is
\begin{equation}\label{eq:QFISLD}
    F_Q(\theta)=\mathrm{tr}\{\rho\, L_{\theta}^2\}.
\end{equation}
The measurement that maximizes the classical Fisher information is a projective measurement in the eigenbasis of the SLD operator $L_{\theta}$. The associated classical Fisher information, computed from these probabilities, achieves the quantum Fisher information. Thus for single parameter estimation, the quantum Cramér–Rao bound can always be saturated \cite{Liu_2019}. 

In quantum sensing with undetected photons, the sample is characterized by two unknown parameters: the transmission $t$ and the phase shift $\phi$. Estimating both simultaneously is therefore a \emph{multiparameter} problem. This setting is intrinsically more subtle than single-parameter estimation: the measurement that is optimal for $t$ need not be optimal for $\phi$, so a single measurement generally cannot saturate both single-parameter bounds at once. Moreover, quantum backaction and measurement incompatibility implies a trade-off: measurement on one parameter can disturb the other.

Suppose we are simultaneously estimating $N$ parameters $\boldsymbol{\theta} = \{\theta_1,\ldots,\theta_N\}$. Then \cref{eq:QFISLD} is generalised to the QFI matrix:
\begin{equation}\label{eq:QFIMatrix}
    \begin{aligned}
        \big[F_Q(\boldsymbol{\theta})\big]_{ij}
        &= \frac{1}{2}\,\mathrm{tr}\!\left[\rho\,(L_{\theta_i}L_{\theta_j} + L_{\theta_j}L_{\theta_i})\right]
         \\
        &= \mathrm{Re}\,[\mathrm{tr}\!\left(\rho\,L_{\theta_i}L_{\theta_j}\right)].
    \end{aligned}
\end{equation}
The Cram\'er-Rao bound in the multiparameter case is then
\begin{gather}
    \sum_j\mathrm{Var}(\hat{\theta}_j)\ge \frac{\mathcal{C}_{\mathcal{S}}}{m},\label{eq:MultiparameterCRBound} \\
    \mathcal{C}_{\mathcal{S}}=\mathrm{tr}\left\{F_Q(\boldsymbol{\theta})^{-1}\right\},
\end{gather}
where $F_Q(\boldsymbol{\theta})^{-1}$ is the inverse of the quantum Fisher information matrix. The term `$\mathcal{C}_{\mathcal{S}}$' stands for `SLD Cram\'er-Rao bound'.

If the SLDs commute
\begin{equation}
[L_{\theta_i},L_{\theta_j}]=0,
\end{equation}
then there exists a single shared eigenbasis, and measuring in this basis will saturate the Cram\'er-Rao bound. In general however the SLDs do not commute. In this case there may not exist a measurement scheme that will attain the bound in \cref{eq:MultiparameterCRBound} \cite{Szczykulska2016,conlon2024roleextendedhilbertspace}. In other words, the bound given by the multiparameter quantum Fisher information is not necessarily tight. To find the fundamental sensing limits in this case, and construct the optimal measurement scheme, we will introduce the Holevo and Nagaoka bounds.

\subsubsection{Holevo bound}

The Holevo bound provides the ultimate precision limit for multiparameter sensing problems considering collective measurements on asymptotically many copies of the probe state. This is defined as \cite{HolevoBook2011} 
\begin{equation}\label{eq:holevo-functional}
    \mathcal{C}_{\mathcal{H}}
    = \min_{\mathbf{X}} \left(\vphantom{\int}\mathrm{Tr}\left\{\mathcal{Z}[\mathbf{X}]\right\} + \mathrm{TrAbs}\left\{\Im\mathcal{Z}[\mathbf{X}]\right\}\right),
\end{equation}
where we minimise over the set $\mathbf{X}=\{X_1,\ldots,X_N\}$ of Hermitian operators satisfying
\begin{align}
    \mathrm{Tr}\{X_i \rho\} &= 0, \label{eq:holevo-constraint1} \\
    \mathrm{Tr}\{X_i \partial_{\theta_j}\rho\} &= \delta_{ij}\label{eq:holevo-constraint2},
\end{align}
and the matrix $\mathcal{Z}$ is
\begin{equation}\label{eq:ZDefinition}
    \mathcal{Z}_{ij}[\mathbf{X}] = \mathrm{Tr}(\rho X_i X_j).
\end{equation}
The second term in \cref{eq:holevo-functional} is defined as 
\begin{equation}
    \mathrm{TrAbs}\{A\} = \mathrm{Tr}\left\{\sqrt{A^\dagger A}\right\}.
\end{equation}
This gives us the \emph{Holevo bound}
\begin{equation}
    \sum_j\mathrm{Var}(\hat{\theta}_j)\ge \frac{\mathcal{C}_{\mathcal{H}}}{m}. \label{eq:HolevoBound}
\end{equation}

The bound in \cref{eq:holevo-functional} consists of two distinct components. The first term in \cref{eq:holevo-functional} acts as an ``SLD-like'' variance contribution, while the second term encodes the skew-information arising from the non-commutativity of the optimal measurements. The gap
\begin{equation}
    \mathcal{C}_{\mathcal{H}} - \mathcal{C}_\mathcal{S} \ge 0
\end{equation}
quantifies the penalty imposed by non-commuting optimal measurements. When the SLDs commute, we have $\mathcal{C}_\mathcal{H}=\mathcal{C}_\mathcal{S}$.

The Holevo bound is attainable, but only in an asymptotic sense. In general we must gather many independent probe states, and perform a collective measurement on them. Then \cref{eq:HolevoBound} gives the information gain per probe state. However, most experiments are limited to sequential measurements of individual probes, or collective measurements on only a few copies of the probe state\,\cite{Conlon_2023_2, Conlon_2023_3}. In particular in \cref{fig:QIUPSetup}, the signal photons are detected individually by photodetectors. To understand the fundamental sensitivity in this case, we must use the Nagaoka bound.

\subsubsection{Nagaoka bound}

Nagaoka's bound characterizes the optimal precision achievable when one is restricted to
separable measurements and classical post-processing \cite{Nagaoka, Conlon_2023, Suzuki_2016}. It therefore quantifies the best performance attainable in realistic measurement scenarios. While the expression when estimating an arbitrary number of parameters is complicated, the Nagaoka bound for two parameters is given by \cite[\S 11]{ Nagaoka} 
\begin{equation}
  \mathcal{C}_{\mathcal{N}}
  = \min_{\mathbf{X}}\left(\;
    \mathrm{Tr}\left\{\mathcal{Z}[\mathbf{X}]\right\}  +
    \mathrm{TrAbs}\left\{
      \rho^{1/2}[X_1,X_2]\rho^{1/2}
    \right\}\right),
  \label{eq:nagaoka-N}
\end{equation}
where as before $\{X_j\}$ are Hermitian operators satisfying \cref{eq:holevo-constraint1,eq:holevo-constraint2}, and $\mathcal{Z}$ is defined as in \cref{eq:ZDefinition}. This gives us the \emph{Nagaoka bound}\footnote{Note that \cref{eq:nagaoka-N} is valid only for the case of two parameters. If there are more than two parameters, we must use the Nagaoka-Hayashi bound \cite{Conlon_2021}.}
\begin{equation}
    \mathrm{Var}(\hat{\theta}_1) + \mathrm{Var}(\hat{\theta}_2)\ge \frac{\mathcal{C}_{\mathcal{N}}}{m}. 
\end{equation}

The first part of \cref{eq:nagaoka-N} is the same as in the Holevo bound. The second term
penalizes the non-commutativity of the locally unbiased estimators $X_i$ and $X_j$
through the sandwiched trace norm. The Nagaoka bound can be saturated using measurements on individual probe states. The optimal POVM is constructed from the eigenvectors of the $X_j$ which minimise \cref{eq:nagaoka-N}. 

In general we have 
\begin{equation}
    \mathcal{C}_{\mathcal{S}} \le \mathcal{C}_{\mathcal{H}} \le \mathcal{C}_{\mathcal{N}}.
\end{equation}
When $\rho$ is pure, the Holevo and Nagaoka bounds coincide \cite{Suzuki_2016}.

\section{Single Parameter Estimation}\label{sec:SingleParameter}
We will first consider estimating either just the transmission, or phase, but not both parameters at once. This corresponds to most current quantum sensing with undetected photon experiments, where only the transmission of the sample is estimated. In this case we can find the limits of single-parameter estimation using the quantum Fisher information.

\subsection{Single pass ultimate sensitivity bound}
The density matrix of a qubit can be parameterised by the Bloch vector $\boldsymbol{r}$ as
\begin{equation}
    \rho = \tfrac12\!\left(I+\boldsymbol{r}\cdot \boldsymbol{\sigma}\right),
\end{equation}
where $\boldsymbol{\sigma}$ is the vector of Pauli matrices. The vector coefficients can be found via $r_j=\mathrm{tr}\{\rho\sigma_j\}$. 
With this, the quantum Fisher information matrix is then given by \cite{Liu_2019}
\begin{equation}\label{eq:QFIBlochVector}
  \bigl[F_Q]_{ij}
  \;=\;
  \partial_{\theta_i}\boldsymbol r . \partial_{\theta_j}\boldsymbol r
  \;+\;
  \frac{(\boldsymbol r\!\cdot\!\partial_{\theta_i}\boldsymbol r)\,(\boldsymbol r\!\cdot\!\partial_{\theta_j}\boldsymbol r)}
       {\,1-|\boldsymbol r|^{2}}.
\end{equation}
When $i=j$, this provides the single-parameter QFI. In our case, the parameters are $\theta_1=t$ and $\theta_2=\phi$. 

For the density matrix of the signal photons in \cref{eq:rho_signal_3state}, the Bloch vector is
\begin{equation}\label{eq:BlochVector}
    \boldsymbol r(t,\phi) = t^{n}\bigl(\cos(n\phi),\, \sin(n\phi),\, 0\bigr).
\end{equation}
The norm is related to the transmission by $\lVert \boldsymbol r \rVert = t^{n}$. 

\begin{figure*}
    \centering
    \includegraphics[width=1\linewidth]{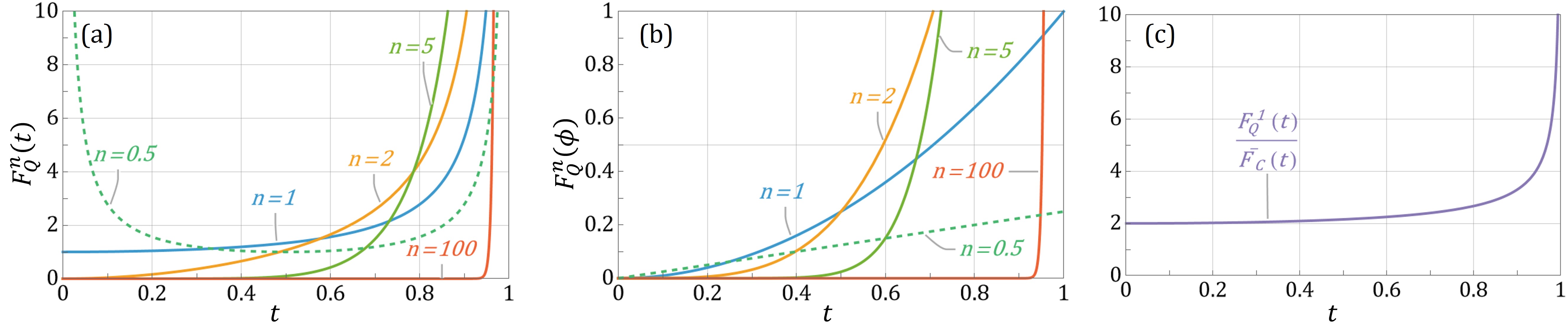}
    \caption{(a) The quantum Fisher information for estimating transmission $t$, given in \cref{eqQFInt}. This is independent of the phase parameter, increases with $t$, and diverges as $t\rightarrow 1$. For $n>1$, increasing the number of passes $n$ decreases the QFI for low transmission samples, and increases it for high transmission. We also consider $n=0.5$, corresponding to halving the path length through the sample, which diverges as $t\rightarrow 0$. As we discuss in \cref{sec:multipass}, this occurs because we are effectively estimating $t$ from measurements of $\sqrt{t}$, and the derivative of the square root is singular at zero. This increases the QFI for low-transmission samples, by increasing the effective transmission. (b) The quantum Fisher information for estimating phase $\phi$, from  \cref{eqFInphi}. This is also independent of $\phi$, but is finite at $t=1$. Again, increasing $n$ is only beneficial for high transmission. Here $n=0.5$ is not singular at $t=0$. This case corresponds to estimating $\phi$ from measurements of $\phi/2$, whose derivative is not singular. For estimation of both transmission and phase, for each $t$ there exists an optimum $n$ (but the same $n$ is not optimal for both transmission and phase estimation). (c) The ratio of the the information provided by the measurement saturating the QFI, to the information provided by the most commonly used measurement in present-day experiments \cref{eq:Fc_per_measurement_discrete}. This assumes a single pass $n=1$. This diverges to infinity as $t\rightarrow 1$, but for any $t < 0.99$, the conventional approach attains a Fisher Information exeeding one tenth of that achievable with the optimal measurement.}
    \label{singpass}
\end{figure*}

Let us first consider the single pass case $n = 1$.  For transmission \cref{eq:QFIBlochVector} gives (see Supplementary Material \S II)
\begin{equation}\label{eq:QFIt}
    F_Q(t)=\frac{1}{1-t^2},
\end{equation}
and for phase
\begin{equation}\label{eq:QFIphi}
    F_Q(\phi)=t^2,
\end{equation}
where we use $F_Q(\theta_j)$ to denote the QFI of the parameter $\theta_j$. We graph these in \cref{singpass} (a) and (b). In both cases, the QFI increases with increasing transmission.

Now let us study the optimal measurement scheme for each parameter. Geometrically, the Bloch vector \cref{eq:BlochVector} is in the equatorial plane, with $t$ controlling the radius, and $\phi$ the azimuthal angle. The two parameters move the Bloch vector in orthogonal directions. Consequently, we will need different measurement schemes for estimation of each parameter.

The optimal measurement bases are the eigenvectors of the symmetric logarithmic derivative with respect to each parameter. To find the SLD for a parameter $\theta$, we write $L_{\theta}=\alpha I+\boldsymbol{\beta}\cdot\boldsymbol{\sigma}$ for unknown coefficients $\alpha,\boldsymbol{\beta}$, and then substitute this into the SLD definition \cref{eq:SLDDefinition}. Detailed calculations are provided in Supplementary Material \S III. For transmission, we find the eigenbasis of the SLD operators to be (again for the $n=1$ case)
\begin{equation}\label{eq:QFItEigenbasis}
    \ket{t_+}
    =\frac{1}{\sqrt2}\!\begin{pmatrix}1\\ e^{i\phi}\end{pmatrix},
    \quad
    \ket{t_-}
    =\frac{1}{\sqrt2}\!\begin{pmatrix}1\\ -e^{i\phi}\end{pmatrix}.
\end{equation}
To confirm that measuring in this basis attains the quantum Fisher information, we can calculate the classical Fisher information for this measurement. For the signal photon state \cref{eq:rho_signal_3state}, the measurement outcome probabilities are
\begin{equation}\label{eq:measuringtps}
    p_\pm = \mathrm{Tr}\left\{\rho\,\lvert t_{\pm}\rangle\langle t_{\pm}\rvert\right\} =\frac{1\pm t}{2}.
\end{equation}
The classical Fisher information $F_C(t)$ is then
\begin{equation}
    F_C(t) = \sum_{\pm}\frac{(\partial_t p_\pm)^2}{p_\pm} = \frac{1}{1-t^{2}} = F_Q(t).
\end{equation}
Thus a projective measurement in the basis in \cref{eq:QFItEigenbasis} is the optimal measurement to estimate the transmission of the sample. We note that since the measurement outcomes are independent of $\phi$, this measurement provides no information about the phase shift.

Now we will consider estimating $\phi$. The eigenvectors of the SLD $L_{\phi}$ are 
\begin{equation}
\ket{\phi_+}
=\frac{1}{\sqrt2}\!\begin{pmatrix}1\\ i\,e^{i \phi}\end{pmatrix},
\quad
\ket{\phi_-}
=\frac{1}{\sqrt2}\!\begin{pmatrix}1\\[2pt] -i\,e^{i \phi} \end{pmatrix}.
\end{equation}
Unlike the case of estimating transmission, the eigenvectors are dependent on the parameter $\phi$ being estimated. QFI measures fluctuations in a parameter about a known operating point. Thus in our analysis, we will let $\phi=\phi_0+\delta\phi$, where $\phi_0$ is assumed known and $\delta\phi$ a small fluctuation. We then measure in the eigenbasis corresponding to $\phi_0$. The outcome probabilities are then 
\begin{equation}\label{eq:ppmphi}
    p_\pm=\frac{1}{2}\left(1 \pm t\,\sin(\delta\phi)\right).
\end{equation} 
At the operating point $\phi=\phi_0$, we have, $p_\pm(\phi_0)=\tfrac12$. The classical Fisher information of this measurement is then
\begin{align}
F_C(\phi)
&=\sum_{\pm}\frac{\big(\partial_{\delta\phi} p_\pm\big)^2}{p_\pm}\Bigg|_{\delta\phi\rightarrow 0}
=t^{2} = F_Q(\phi).
\end{align}
This is equal to the SLD quantum Fisher information, confirming that this measurement is optimal. Since when $\delta\phi=0$ the measurement probabilities \cref{eq:ppmphi} are $p_{\pm}=1/2$, this measurement provides no information about the transmission.

A key observation is that the eigenvectors of the SLD operators depend explicitly on the \emph{unknown} phase $\phi$. Therefore, the projective measurement that attains the quantum Cram\'er--Rao bound is inherently \emph{local}: it is optimal only in a neighbourhood of some reference value~$\phi_0$. Practically, this means that we must first obtain a coarse phase estimate $\hat{\phi}=\phi_0$, and then choose a measurement basis that is aligned with the SLD eigenvectors at $\phi \simeq \phi_0$. 

Let us now discuss how to implement these measurements. Both projective measurements can be implemented via the controllable phase shift $\theta$ in the Measurement block in \cref{fig:QIUPSetup}. For the transmission measurement we have $\theta=\phi_0$, while for the phase measurement we have $\theta=\phi_0+\pi/2$.

A second observation is that the quantum Fisher information (QFI) for both parameters, $t$ and $\phi$, depend only on the transmission $t$. In particular, sensitivity increases with $t$. However, even though the QFI lacks explicit $\phi$-dependence, the \emph{optimal measurement} does depend on approximate knowledge of $\phi$. These results motivate a two-step protocol: first a coarse phase acquisition, followed by measurement in the optimal basis.

In practice the sample-induced phase shift is expected to be small, so we may assume $\phi$ lies in a neighbourhood of zero. We therefore begin with a short \emph{phase-acquisition} step: we scan the controllable phase $\theta$ over a few settings around $0$ and fit the resulting interference response to obtain a coarse estimate $\hat{\phi}$. This rough estimate is sufficient to choose a suitable reference phase $\phi_0 \approx \hat{\phi}$ (or more generally to pick $\phi_0$ that maximises the Fisher information at the operating point), after which we can perform the final measurement in the corresponding optimal basis.

\subsection{Comparison with existing experiments}
\label{subsec:exp_est_t}

In this section we will consider the measurement scheme used in current experiments for sensing with undetected photons. These estimate only the transmission $t$ of the sample. We perform $M$ independent measurements, sweeping the value of $\theta$ in the measurement block of \cref{fig:QIUPSetup} between $0$ and $2\pi$:\footnote{In practice this is achieved by varying the path length of one of the signal arms.}
\begin{equation}
    \theta=\left\{0,\frac{2\pi}{M},2\times\frac{2\pi}{M}\dots,2\pi\right\}.
\end{equation}
The detection probabilities at each photodetector will vary sinusoidally, with visibility $t$. The transmission is thus estimated by fitting this sine wave. We will now compare the information gain from this measurement, to that of the optimal measurements.

Let us consider the probability distribution for the $k$th measurement. This has two outcomes $x_k\in\{+1,-1\}$, representing a photon click at either of the two detectors. Following \cref{eq:ModelDetectionProbabilities} the probability of each outcome is (again assuming the single-pass case $n=1$)
\begin{equation}
    p_k(x_k) = \frac12\!\left(1 + x_k\, t \cos\left(\phi+k\frac{2\pi}{M}\right)\right).
\label{eq:pxk_t_phi_chik}
\end{equation}
Each of the measurements is independent of the others. The probability of a sequence of measurement results $\mathbf{x}=\{x_1,\ldots,x_M)$ is then the product of the probabilities:
\begin{equation}
    P(\mathbf{x})=\prod_{k=1}^{M} p_k(x_k).
\label{eq:joint_likelihood}
\end{equation}

\begin{figure*}
    \centering
    \includegraphics[width=1\linewidth]{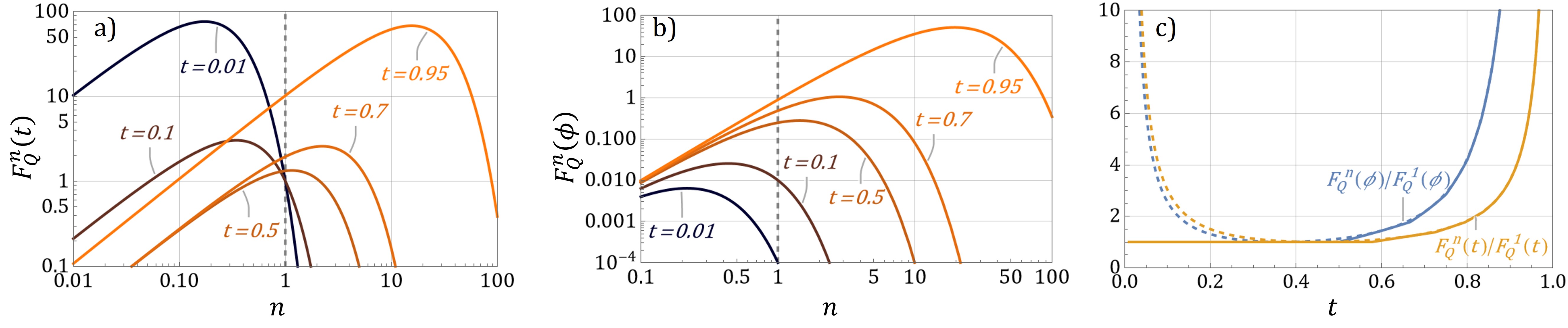}
    \caption{(a) The quantum Fisher information for estimating transmission (\cref{eqQFInt}), as a function of number of passes $n$. For each transmission, there is an optimum $n$ which maximises the QFI. In general, increasing $t$ always leads to higher QFI. However when $n<1$, corresponding to shortening the path length through the sample, low transmission samples also have a higher QFI because the shortened path length corresponds to a higher effective transmission. (b) The quantum Fisher information for estimating phase shift (\cref{eqFInphi}). Again there is an optimum number of passes for each transmission value, which is different to the optimum $n$ for estimating $t$. Here, increasing $t$ always leads to an increase in QFI. (c) To quantify the enhancement provided by multipass setups, we plot the ratio of QFIs for optimal $n$, to the information provided by a single-pass setup. For the solid curves, the optimal $n$ is rounded to the nearest integer, representing a true multipass setup. The dashed curves allow for non-integer $n$, corresponding to varying the path length through the sample. Dashed and solid curves overlap for high transmission. For low transmission we see an enhancement when path length is shortened. For moderate transmissions, the possible enhancement provided by multipass setups is less than an order of magnitude. However the advantage significantly increases as $t\rightarrow 1$.}
    \label{fig:phi}
\end{figure*}

The Fisher information of a product of probability distributions, is equal to the sum of Fisher informations of the distributions. Thus the Fisher information of $P(\mathbf{x})$ with respect to $t$ is
\begin{equation}\label{eq:Fc_single_def}
    \begin{aligned}
    F_C\left(t\right)
     &= \sum_{k=1}\left(\sum_{x_k\in\{+,-\}} \frac{\left(\partial_t p_k(x_k)\right)^2}{p_k(x_k)}. \right) \\
     &= \sum_{k=1}^M\frac{\cos^2\left(\phi+k\frac{2\pi}{M}\right)}{1-t^2\cos^2\left(\phi+k\frac{2\pi}{M}\right)}.
    \end{aligned}
\end{equation}
Typically the number of measurements $M$ is large, so let us take the continuum limit. Defining $\Delta\theta=2\pi/M$, we can re-write the above as
\begin{equation}
    F_C=\frac{1}{\Delta\theta}\left(\sum_{k=1}^M \frac{\cos^2\left(\phi+k\Delta\theta\right)}{1-t^2\cos^2\left(\phi+k\Delta\theta\right)}\Delta\theta\right).
\end{equation}
To compare with the optimal measurement schemes we found, we must compute the Fisher information per measurement: $F_C/M$. Taking the limit as $M\rightarrow \infty$, and hence $\Delta\theta\rightarrow 0$, we find
\begin{equation}
    \begin{aligned}
        \bar{F}_C(t) &= \lim_{M\rightarrow\infty}\frac{F_C(t)}{M}, \\
        &= \frac{1}{2\pi}\int_0^{2\pi}\frac{\cos^2\left(\phi+\theta\right)}{1-t^2\cos^2\left(\phi+\theta\right)}\mathrm{d}\theta,
    \end{aligned}
\end{equation}
where now $\bar{F}_C$ is the Fisher information per measurement, in the limit that number of measurements $M$ approaches infinity. Since the integrand is periodic, the integration value is independent of $\phi$. We can thus evaluate this at $\phi=0$ as
\begin{equation}\label{eq:Fc_per_measurement_discrete}
    \bar F_C(t)= \frac{1-\sqrt{1-t^2}}{t^2\sqrt{1-t^2}}.
\end{equation}

In \cref{singpass}(c) we compare this to the optimal $F_Q(t)$ in equation \cref{eq:QFIt}. We find that the ratio $F_Q(t)/\bar{F}_C(t)$ is strictly $>1$. At low transmissions the optimal measurement scheme doubles the available information. The ratio diverges to infinity as $t\rightarrow 1$.
Hence, having (even approximately) the correct phase reference enables strictly higher Fisher information. This is because the optimal measurement aligns with the direction that changes the most with the parameter. The phase-scanning measurement however measures at a range of phase values, only one of which is providing the optimum amount of information.

\subsection{Multi-pass ultimate sensitivity bound}\label{sec:multipass}
We can enhance the signal by passing the photon multiple times through the sample, as shown in the `Multipass Interaction' box in \cref{fig:QIUPSetup}. This increases the effective phase shift and absorption of the idler photon. Two-pass setups are used in folded configurations used in microscopy \,\cite{paterova2017nonlinear, paterova2018tunable}. However, there is a tradeoff when using multiple passes. While the effect of the parameters on the probe is magnified, at the same time our signal is reduced by absorption of photons by the sample. In this section we will study the optimum number of passes, and find the ultimate limits on precision enhancement from multiple passes. 

In the case of multiple passes, the quantum Fisher informations become (see Supplementary Material \S II)
\begin{align}
    F_Q^n(t) &= \frac{n^2}{t^2}\frac{t^{2n}}{1-t^{2n}},\label{eqQFInt} \\
    F_Q^n(\phi) &= n^{2}t^{2n}. \label{eqFInphi}
\end{align}
We note that this differs from substituting $t\rightarrow t^n,e^{i\phi}\rightarrow e^{in\phi}$ in the single-pass expressions \cref{eq:QFIt,eq:QFIphi}. In \cref{singpass}(a) and (b) we plot $F_Q^{(n)}(t)$ and
$F_Q^{(n)}(\phi)$ for $n\in\{1,2,5, 100\}$. These show that for a given transparency, there is an optimum number of passes, above or below which information is degraded. In particular, for low transparency, multiple passes can severely degrade the signal.

In \cref{fig:phi} (a) and (b) we plot the QFI for fixed $t$, versus pass number $n$. As transmission approaches 1, the optimum $n$ grows larger. Thus a large number of passes provides a benefit primarily for high transmissions. We also consider values of $n$ below one, corresponding to decreasing the path length of the idler mode through the sample. For phase estimation, regardless of the value of $n$, increasing transmission always increases the QFI. Transmission estimation follows the same trend for $n>1$. However for $n<1$, high QFI occurs for both low and high $t$. We will now explore this phenomenon.

In \cref{singpass} (a), we see that the quantum Fisher information for $n=0.5$ is singular as $t\rightarrow 0$, and qualitatively different from the $n>1$ cases. To understand this, let us consider the optimal measurement signal \cref{eq:measuringtps} in the multipass case:
\begin{equation}\label{eq:measuringtmp}
    p_{\pm}=\frac{1\pm t^n}{2}.
\end{equation}

Effectively, we are measuring $t^n$, in order to estimate $t$. The variation in our measurement probabilities is
\begin{equation}\label{eq:dpt}
    \partial_tp_{\pm}=\pm \frac{n}{2}t^{n-1}.
\end{equation}

Let us consider this in the limit $t\rightarrow 0$. When $n>1$, \cref{eq:dpt} is always zero in this limit. Thus our measurement signal is to first order insensitive to small increases in $t$ from zero. When $n<1$ (corresponding to reducing the path length through the sample), \cref{eq:dpt} is infinite, which is the reason for the singular behaviour in \cref{singpass} (a) for $n=0.5$. It is only in the $n=1$ case that \cref{eq:dpt} is neither zero nor infinity. For phase estimation in \cref{singpass} (b) however, our measurement signal \cref{eq:ppmphi} in the multipass case is
\begin{equation}
    p_{\pm}=\frac{1\pm t^n\sin(n\delta\phi)}{2}.
\end{equation}
The derivative of this function with respect to $\delta\phi$ is zero as $t\rightarrow 0$ regardless of $n$, hence the non-singular behaviour.

We investigate this further in the appendices. In \cref{app:SimulatingDeltatErr} we simulate estimating transmission. We find that we do indeed observe an increase in estimation accuracy of $t$ at low transmissions when we take $n=0.5$, and a collapse in sensitivity for $n=2$. 
The transmission and phase shift can also be written in terms of material parameters as $t=e^{-\gamma L}$ and $\phi=\kappa L$, where $L$ is the path length of the idler through the sample, and $\gamma,\kappa$ are the material absorption and phase shift per unit length. In \cref{app:MaterialCoefficients} we calculate the QFI for estimation of $\gamma$ and $\kappa$. With this reparameterisation, the QFI is no longer singular as $t\rightarrow 0$. We find that the QFIs of these parameters behave similarly to one-another, and do not show any change in behaviour for $L<1$. 

Whether we are more interested in the net transmission and phase shift, or the material parameters, depends on the situation. For imaging, where we may be looking at a biological sample containing many different components, the information we are interested in is the overall transmission and phase shift. Thus we will continue to focus on these parameters for the rest of the manuscript.

In order to find the ultimate sensitivity limits, we consider the case where $n$ is controllable, and can be tuned to optimise the measurement. For example we may have an automated multipass cell, or the sample may be gasseous, in a chamber of variable length. Let us first consider estimating the sample transmission. We seek the value $n^*_t$ that maximizes $F^n_Q(t)$ for a given $t$. Setting the derivative $\partial_nF_Q^n(t)=0$, we find the relation
\begin{equation}\label{eq:optnt}
    1+n^*_t\ln t=t^{2n^*_t}.
\end{equation}
This has a unique solution for $n^*_t$, which can be found numerically. It seems reasonable that the multipass interaction seeks to create an effective optimal transparency:
\begin{equation}
    t^*=t^{n_t^*}.
\end{equation}
Substituting this ansatz into \cref{eq:optnt} gives the relation
\begin{equation}
    1+\ln t^*-(t^*)^2=0,
\end{equation}
which numerically has solution $t^*\approx 0.45$. Thus the optimum number of passes is
\begin{equation}\label{eq:optnt2}
    n_t^*\approx \frac{\ln 0.45}{\ln t}\approx\frac{-0.8}{\ln t}.
\end{equation}

Now suppose we want to estimate the phase. Again setting $\partial_nF_Q^n(\phi)=0$, we find the optimum number of passes $n_{\phi}^*$ to be
\begin{align}\label{eq:optnphi}
    n_{\phi}^*=\frac{-1}{\ln t},
\end{align}
corresponding to $t^{n_{\phi}^*}=1/e$. Thus, the optimal number of passes for transmission, is different from that of the optimal number for estimating phase. In both cases though they scale as $n^*\sim 1/\lvert\ln t\rvert$.

In \cref{fig:phi}(c) we plot the ratio between the quantum Fisher information evaluated at the optimal number of passes, compared to the single-pass setup. When $n$ is restricted to integers, multipass setups only deliver advantage when transmission is greater than half. For highly transparent samples, the multipass strategy can provide significant enhancement. For a sample with $t=0.99$, the optimal configuration for phase estimation uses $n_{\phi}^*=100$ passes to increase the QFI by a factor of 1400 compared to the $n=1$ case. However, for transmission estimation we have $n_t^*=79$, increasing the QFI by a factor of approximately 30. For $t=0.999$, we have $n_t^*\approx 800$ and $n_{\phi}^*\approx 1000$, enhancing the QFI by $~300$ times for transmission and $130,000$ times for phase. Thus multipass schemes are particularly useful for precision sensing of the phase shift of a sample.


\section{Multiparameter Estimation}\label{sec:MultiparameterEstimation}
In general, both the transmission and phase shift of the sample must be estimated simultaneously. For this multiparameter case, we can consider the quantum Fisher information matrix given in \cref{eq:QFIMatrix}:
\begin{equation}
    \mathcal{F}_Q(t,\phi)=
    \begin{pmatrix}
        \dfrac{n^{2} t^{\,2n-2}}{1-t^{\,2n}} & 0\\[6pt]
        0 & n^{2} t^{\,2n}
    \end{pmatrix}.
\end{equation}
The corresponding multiparameter quantum Cram\'er--Rao bound is obtained from the trace of the inverse matrix \cref{eq:MultiparameterCRBound}:
\begin{equation}\label{eq:MultiparameterBoundCalculated}
    \mathcal{C}_S
    = \Tr\!\bigl[\mathcal{F}_Q^{-1}\bigr]
    = \frac{1-t^{2n}}{n^{2} t^{\,2n-2}}
      + \frac{1}{n^{2} t^{\,2n}}.
\end{equation}
This gives a lower bound on the sum of the variances of any locally unbiased estimators of $t$ and $\phi$. However, the SLD operators $L_t$ and $L_\phi$ do not commute. Consequently, it does not necessarily follow that there is a single projective measurement that can attain \cref{eq:MultiparameterBoundCalculated}. To characterise the \emph{achievable} performance, we must use the more sophisticated machinery of the Holevo and Nagaoka bounds.

\subsection{Holevo and Nagaoka bounds}
 
The Holevo and Nagaoka bounds are given by optimisation problems over Hermitian operators $X_t$ and $X_{\phi}$ satisfying \cref{eq:holevo-constraint1,eq:holevo-constraint2}. 
These act on the same Hilbert space as the density matrix. Thus we can parameterise them as two-by-two matrices with real diagonal entries and complex-conjugate off-diagonal entries:
\begin{align}\label{eq:Xt}
    X_t &=
    \begin{pmatrix}
        a & b+i c\\
        b-i c & d
    \end{pmatrix}, \\
    X_{\phi} &=\label{eq:Xphi}
    \begin{pmatrix}
        e & f+i g\\
        f-i g & h
    \end{pmatrix}.
\end{align}
Here we have eight unknown parameters $a,b\ldots,h$. The conditions  \cref{eq:holevo-constraint1,eq:holevo-constraint2} provide six constraints, leaving us with two free parameters. For details on the calculations in this section, refer to Supplementary Material \S IV.

Let us first consider the Holevo bound. Substituting \cref{eq:Xt,eq:Xphi} into the optimisation problem \cref{eq:holevo-functional} gives a quadratic equation in the two free parameters, which can be minimised by algebraic manipulation. This gives the bound
\begin{equation}
    \mathcal{C}_{\mathcal{H}}
    = \frac{1-t^{2n}}{n^{2} t^{\,2n-2}}
      + \frac{1}{n^{2} t^{\,2n}}
    = \mathcal{C}_S.
\end{equation}
Thus the Holevo bound coincides with the multiparameter quantum Cram\'er-Rao bound. The Holevo bound can be achieved asymptotically, representing the information per signal photon given by a joint measurement on multiple signal photons at once. However, it cannot in general be attained by measurements on individual signal photons, as is done in most imaging setups such as \cref{fig:QIUPSetup}.

\begin{figure*}
    \centering
    \includegraphics[width=1\linewidth]{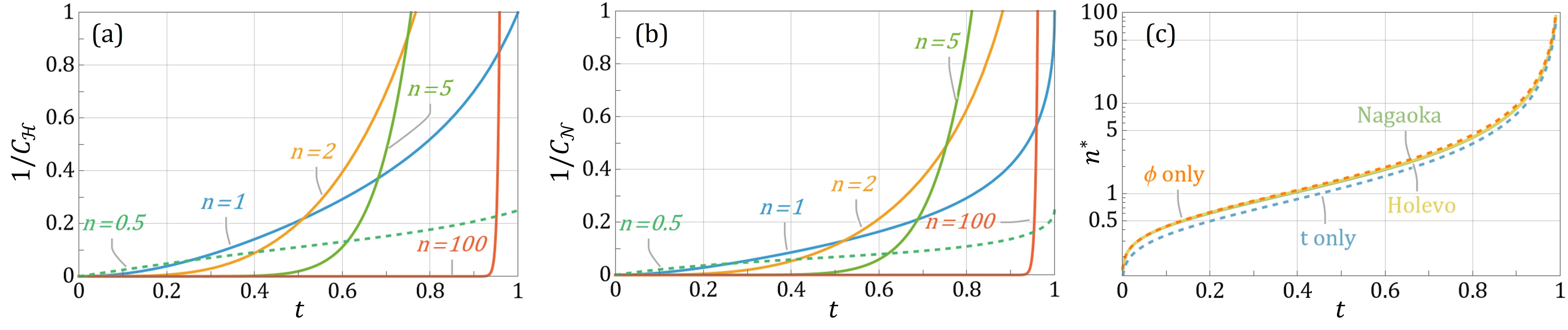}
    \caption{(a) and (b) plot the available two-parameter information according to the Holevo and Nagaoka bounds, as a function of $t$. Both show similar qualitative behaviour to each other, and the quantum Fisher information in \cref{singpass}. We can see that the information from the Holevo bound is approximately 25\% larger, corresponding to additional quantum advantage which could be accessed using joint measurements on multiple probes. (c) The optimal number of passes, which maximise the Holevo, Nagaoka, and quantum Fisher informations. Curves for the Nagaoka and Holevo bounds overlap. We see that all obey the same general trend. For samples with transmission below $0.8$, optimum information can be achieved with fewer than four passes. As $t\rightarrow 1$ the optimum number of passes grows towards infinity.
    }
  \label{fig:widepair2}
\end{figure*}

To understand the limit for schemes involving measurement of individual signal photons, we use the Nagaoka bound. We begin by substituting \cref{eq:Xt,eq:Xphi} into the optimisation in \cref{eq:nagaoka-N}. The first term $\mathrm{Tr}\{\mathcal{Z}[\mathbf{X}]\}$ is a quadratic equation, which can easily be minimised. The second term $\mathrm{TrAbs}\{\rho^{1/2}[X_t,X_{\phi}]\rho^{1/2}\}$ is complex, but can also be manipulated into a form which can be minimised. We find that the same $X_t,X_{\phi}$ minimise both terms. With this, we find the Nagaoka bound:
\begin{equation}
    \mathcal{C}_{\mathcal{N}} = \frac{\bigl(1+t\sqrt{\,1-t^{\,2n}\,}\bigr)^{2}}{n^{2}\,t^{\,2n}}.
\end{equation}
We can verify that $\mathcal{C}_{\mathcal{N}}>\mathcal{C}_{\mathcal{H}}=\mathcal{C}_{\mathcal{S}}$. Thus the bound given by the quantum Fisher information is only attainable through collective measurements on multiple probes.

As in the single-parameter setting, all three bounds depend only on the transmission $t$, and are independent of the phase shift $\phi$. We plot them in \cref{fig:widepair2}(a) and (b). We can see that both bounds have the same qualitative behaviour. As $t \to 0$, the inverses of both the Holevo and Nagaoka bounds vanish. This is because these bounds are proportional to the sum of estimation variances of $t$ and $\phi$. As $t \to 0$, the variance of estimate of $\phi$ diverges, causing the total variance to diverge. In the limit $t\rightarrow 1$, the information in both cases grows to infinity.  As in the single-parameter case, increasing the number of passes increases the information for high transparencies, and decreases it for low transparencies.

\subsection{Measurement saturating the Nagaoka bound}
Let us now construct the optimal measurement scheme, which can attain the Nagaoka bound. This can be constructed from the eigenvectors of the $X_t,X_{\phi}$ which minimise \cref{eq:nagaoka-N} \cite{Conlon_2021, Conlon_2023}. As shown in Supplementary Material \S IV, these are
\begin{align}
    X_t &=
    \frac{t}{n}\begin{pmatrix}
        -1 & t^{-n}e^{-i n \phi}\\
        t^{-n} e^{i n \phi} & -1
    \end{pmatrix}, \\
    X_{\phi} &=
    \frac{i}{nt^n}\begin{pmatrix}
        0 & e^{-i n \phi}\\
        -e^{i n \phi} & 0
    \end{pmatrix}.
\end{align}
These have normalised eigenvectors
\begin{align}
    \ket{X_{t+}} &=
    \frac{1}{\sqrt{2}}\begin{pmatrix}
        e^{i n \phi}\\
        1
    \end{pmatrix},
    &
    \ket{X_{t-}} &=
    \frac{1}{\sqrt{2}}\begin{pmatrix}
        -e^{i n \phi}\\
        1
    \end{pmatrix}, \label{eq:XtEigenvectors}\\
    \ket{X_{\phi+}} &=
    \frac{1}{\sqrt{2}}\begin{pmatrix}
        i e^{i n \phi}\\
        1
    \end{pmatrix},
    &
    \ket{X_{\phi-}} &=
    \frac{1}{\sqrt{2}}\begin{pmatrix}
        -i e^{i n \phi} \\
        1
    \end{pmatrix},\label{eq:XphiEigenvectors}
\end{align}
in terms of which we define the projectors
\begin{align}
    \Pi(X_{t+}) &= \left\lvert X_{t+}\right\rangle\left\langle X_{t+}\right\rvert, \\
    \Pi(X_{t-}) &= \left\lvert X_{t-}\right\rangle\left\langle X_{t-}\right\rvert, \\
    \Pi(X_{\phi+}) &= \left\lvert X_{\phi+}\right\rangle\left\langle X_{\phi+}\right\rvert, \\
    \Pi(X_{\phi-}) &= \left\lvert X_{\phi-}\right\rangle\left\langle X_{\phi-}\right\rvert.
\end{align}
Let $0<\lambda<1$, and define the POVM
\begin{equation}\label{eq:NagaokaPOVM}
    \begin{aligned}
        &\left\{\lambda\,\Pi(X_{t+}),\;\lambda\,\Pi(X_{t-}), \right.\\
        &\hspace{1em}\left.(1-\lambda)\,\Pi(X_{\phi+}),\; (1-\lambda)\,\Pi(X_{\phi-})\right\}.
    \end{aligned}
\end{equation}
This corresponds to a measurement scheme where with probability $\lambda$ we randomly switch between projective measurement in the eigenbases of $X_t$ and $X_{\phi}$.

To find the optimum value of $\lambda$, we calculate the variance of this measurement. The minimum variance occurs at
\begin{equation}\label{eq:NagaokaLambdaStar}
    \lambda^{*}
    =\frac{t\sqrt{1-t^{2n}}}{1+t\sqrt{1-t^{2n}}}.
\end{equation}
At this point, we calculate that measurement in the POVM has equality with the Nagaoka bound:
\begin{equation}
    \mathrm{Var}(\lambda^*)=\mathcal{C}_{\mathcal{N}}.
\end{equation}
Thus this is the optimal measurement. For any $\lambda\neq\lambda^*$ the variance is strictly above $\mathcal{C}_{\mathcal{N}}$, and the measurement is suboptimal.

Now let us consider the physical implementation, in the `Measurement' block of \cref{fig:QIUPSetup}. Prior to the beamsplitter, we can regard $|0\rangle$ as the path containing the phase shift $e^{i\theta}$, and $|1\rangle$ as the path without a phase shift. Then from \cref{eq:XtEigenvectors}, a photodetection with $\theta=n\phi$ corresponds to a projective measurement in the basis $\{\Pi(X_{t+}),\Pi(X_{t-})\}$. Similarly from \cref{eq:XphiEigenvectors}, photodetection with $\theta=n\phi + \pi/2$ corresponds to projective measurement in the basis $\{\Pi(X_{\phi+}),\Pi(X_{\phi-})\}$. The POVM \cref{eq:NagaokaPOVM} can be realised by, for each incoming signal photon, choosing $\theta=n\phi$ with probability $\lambda^*$, or $\theta=-n\phi$ with probability $1-\lambda^*$. Equivalently if we are measuring $N$ signal photons, we can measure the first $\lambda^*N$ using $\theta=n\phi$, and the rest using $\theta=-n\phi$. We note that the eigenbases of $X_t$ and $X_{\phi}$ are the  optimal measurement bases for transmission and phase shift respectively, provided by the quantum Fisher information. 

In \cref{fig:widepair2}(c) we plot the optimal number of passes that optimise the Holevo and Nagaoka bounds, as well as the quantum Fisher information for each of the parameters \cite{onlinecode}. While for the QFI the optimum $n_t^*$ satisfies a simple relation $t^{n_t^*}\approx 0.45$, we find that no such simple relationship exists for the Holevo and Nagaoka bounds. However, the optimal $n$ is approximately the same for each of the bounds.

\subsection{Optimising the number of passes for each parameter}
With the quantum Fisher information, we found optimal measurement schemes for estimating either transmission or phase. However, each scheme was suitable for estimating only one parameter. Using the Nagaoka bound, we then showed that dividing our signal photons between these schemes with ratio \cref{eq:NagaokaLambdaStar} provides the optimal measurement for simultaneous estimation of both parameters. In this analysis, both schemes used the same number of passes $n$, which is a reasonable experimental constraint. In principle however, we could use some actuator to vary the number $n$ of passes used by each scheme. In this section we will analyse this case, to find the ultimate limit to quantum sensing with undetected photons.

Let us denote the total number of signal photons we will measure as $N$. For some $0\le x\le 1$ we allocate 
\begin{equation}
    N_t=\lceil xN\rfloor
\end{equation}
(where $\lceil xN\rfloor$ denotes $xN$ rounded to the nearest integer) to the optimal measurement scheme for $t$, and 
\begin{equation}
    N_{\phi}=N-N_t
\end{equation}
to the optimal measurement scheme for $\phi$. If we must choose the same number of passes $n$ for both schemes, then the optimum $x$ is given by $\lambda^*$ in \cref{eq:NagaokaLambdaStar}. However, now we will let the number of passes for the transmission measurement be $n=n_t^*$ as in \cref{eq:optnt}, and for the phase measurement we take $n=n_{\phi}^*$ as in \cref{eq:optnphi}.

From the Cram\'er-Rao bound, the sum of variances in this mixture strategy will be lower bounded by
\begin{equation}\label{eq:mixturevariance}
    \mathrm{Var}(\hat{t})+\mathrm{Var}(\hat{\phi})\ge\frac{1}{N_t}\frac{1}{F_Q^{n^*_t}(t)}+\frac{1}{N_{\phi}}\frac{1}{F_Q^{n^*_{\phi}}(\phi)},
\end{equation}
where the terms on the denominator are the Fisher informations \cref{eqQFInt,eqFInphi}. This can be saturated using the measurement schemes discussed in \cref{sec:SingleParameter}. We then numerically search for the $x$ minimising the right-hand-side of \cref{eq:mixturevariance}.

\begin{figure}
    \centering
    \includegraphics[width=1\linewidth]{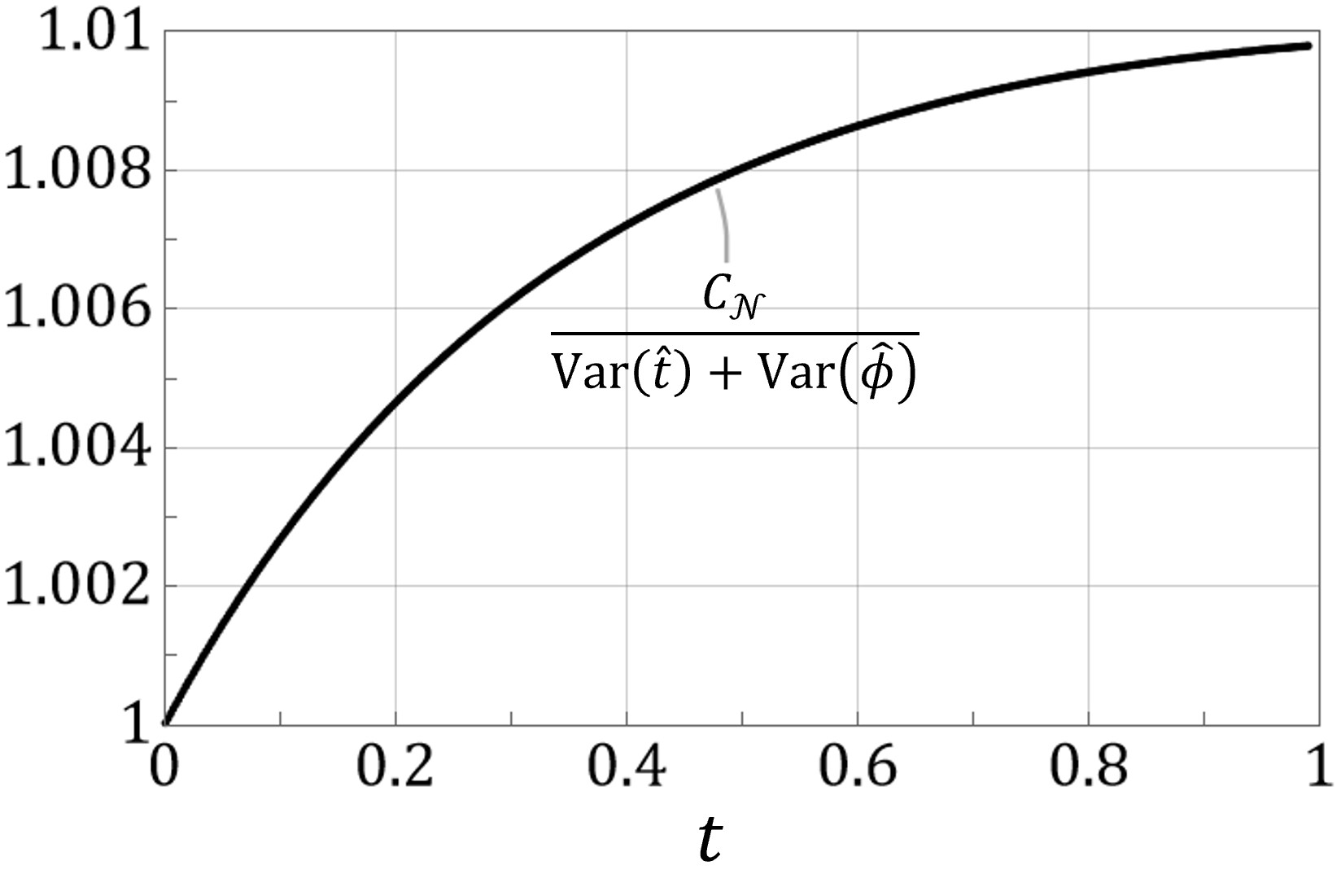}
    \caption{The ratio of the variance given by the Nagaoka bound (which uses the same number of passes for each parameter), to the variance in \cref{eq:mixturevariance}. The total number of measurements $N$ is taken to be infinite. Using different $n$ provides an improvement which increases as transparency approaches $1$, peaking at $1\%$.
}
  \label{figjoint}
\end{figure}

We plot in \cref{figjoint} the ratio of the variance of the measurement with different $n$s, to the variance given by the Nagaoka bound. Here $N$ is taken to be infinity, removing effects of discretisation when rounding $\lfloor xN\rceil$.\footnote{In most experiments $N$ is large enough that discretisation is irrelevant.} We see that the advantage of choosing a different path length for measurements of $t$ and $\phi$ increases as transmission increases, but peaks at only $1\%$.\footnote{The reader may ask how it is possible to measure with precision surpassing what is given by the Nagaoka bound. The reason is that Nagaoka's bound supposes that the same measurement scheme is used to measure both parameters. Here, by using two different $n$, we are effectively combining results from two different measurement schemes. Hence Nagaoka's bound does not apply.} Thus this result shows that the measurement saturating the Nagaoka bound is very close to optimal. This is because the optimum values of $n$ for either parameter \cref{eq:optnt2,eq:optnphi}, while not equal, are similar. Thus there is limited advantage to be gained in tuning $n$ separately for each parameter.

\section{Discussion}

We established fundamental limits for simultaneously estimating the transmission $t$ and phase shift $\phi$ of a sample in quantum sensing with undetected photons. We showed that the optimal measurement scheme requires only a controllable phase shift. A key requirement is an initial estimate of the sample phase shift, which is used to construct the optimal measurement basis. This is a common scenario in precision sensing, where we seek to detect infinitesimal fluctuations about a known operating point. Examples of this include detecting the appearance of traces of an unknown gas, or monitoring for slight changes in a cell or biological tissue over time. We can also begin by performing a coarse estimate of the phase using standard methods. 

In many cases however, such as imaging an unknown object, the requirement of known initial phase may be too strict. Our results in \cref{singpass}(c) show that the current widely-used experimental strategy, which does not require a priori phase estimate, performs within an order of magnitude of the optimum strategy unless the sample is highly transmissive. It would be of interest for future work to find the optimum strategy when both the phase and transmission are completely unknown. Bayesian methods can treat this case, allowing for $t$ and $\phi$ to be random variables with prior distributions  \cite{Demkowicz2015,suzuki2023bayesiannagaokahayashiboundmultiparameter}. Bayesian methods can also analyse adaptive protocols, which could provide better performance by continually optimising the measurement over time.

We were also able to quantify the advantage provided by multipass configurations. These are a natural avenue for enhancing a weak interaction with a sample, however there is a trade-off due to signal degradation from absorption of idler photons. We identified simple formulae for the optimum number of passes, which scale as $n^*\sim 1/\lvert\ln t\rvert$. It suffices to choose a single optimum number of passes to measure both parameters. For samples with transmission less than 0.8 the optimum number of passes is less than four. Highly transmissive samples can benefit strongly from setups with a large number of passes. However, this must be balanced with the experimental difficulty of mode-matching the idler mode and maintaining the coherence conditions over these passes. Moreover, increasing the Fisher information by a factor, gives the same sensitivity enhancement as increasing the pump power by the same amount. One notable result was that for very low transmission, two passes significantly degrades the sensitivity compared to a single pass. This is relevant to many many microscopy setups, which use an effective two-pass interaction with a folded interferometer. 

While the present analysis relies on interference at individual photodetectors, many applications of sensing with induced coherence are in multi-pixel imaging. Transitioning to an imaging configuration introduces specific challenges for the measurement scheme. We have shown that the optimal measurement scheme requires approximate knowledge of the phase $\phi$. In an imaging scenario, the phase $\phi$ will generally vary at each pixel. To resolve this we can introduce a spatial light modulator (SLM) before the detector, which can apply such local phase shifts. Alternatively we could apply a global phase shift but sample only the pixels for which that phase shift is optimal. Another approach is to simply raster scan each pixel. A promising direction for future research would be to explicitly study the imaging problem, and study how an SLM could be used to perform optimal measurements beginning from a totally unknown phase.

All current experiments measure signal photons one at a time. However, the Holevo bound shows that if multiple signal photons can be measured at once, simultaneously using collective measurements, a further quantum enhancement is achievable. Although the Holevo bound can never be achieved exactly\,\cite{conlon2024gappersistencetheoremquantum}, it would be interesting to explore collective measurement schemes which approach this bound. Moreover, quantum resources such as squeezed light may further improve sensitivity, especially for phase estimation, potentially shifting the multiparameter trade-offs. Finally, it would be enlightening to  incorporate imperfections such as external loss, detector inefficiency, phase noise, and mode mismatch. Such analysis would allow us to design experimental schemes which are robust to noise.

\section{Acknowledgements}
We are grateful for discussions with Thomas Produit, Leonid Krivitsky, Anna Paterova, Tanmoy Chakraborty, Aritra Das,  Aaron Tranter, and Chenyue Gu. \cref{fig:QIUPSetup} was made using the open source software Blender, and quantum optics assets created by Ryo Mizuta Graphics \cite{mizuta_optical_components_v1}. This work is  supported by the National Research Foundation, Singa
pore through the National Quantum Office, hosted in
A*STAR, under its Centre for Quantum Technologies Funding Initiative (S24Q2d0009), as well as the A*STAR Quantum Innovation Centre (Q.InC) SRTT. Lorc\'{a}n O. Conlon was supported by NSF QLCI (award No.~OMA-2120757), and the Templeton Foundation grant 63121. 

\bibliography{ref}

\clearpage

\appendix

\section{Monte Carlo study of estimating transmission}\label{app:SimulatingDeltatErr}

In this section we present some simulations to support the results in \cref{sec:multipass}, showing that the QFI can be enhanced for low-transmission samples by taking $n<1$. For the optimal measurement scheme for estimation of transmission, the outcome probabilities are given by \cref{eq:measuringtmp}. The QFI quantifies our ability to detect a small fluctuation $\Delta t$ in $t$. Thus we consider
\begin{equation}\label{eq:DetectingtDelta}
    p_{\pm}(t+\Delta t)=\frac{1\pm (t+\Delta t)^{n}}{2}.
\end{equation}
We can invert this to find:
\begin{equation}
    \Delta t = \left(p_+-p_-\right)^{1/n}-t.
\end{equation}
This allows us to estimate $\Delta t$, from an estimate of the measurement probabilities.

Suppose we perform $N$ photodetections in sequence. Let $n_{+}$ be the number of ``$+$'' outcomes, and $n_{-}$ the number of ``$-$'' outcomes. Then we can estimate the corresponding probabilities as
\begin{equation}
    \begin{aligned}
        \hat{p}_+ &= \frac{n_+}{N}, \\
        \hat{p}_- &= 1-\hat{p}_{+}.
    \end{aligned}
\end{equation}
This gives us our estimator for $\Delta t$:
\begin{equation}\label{eq:DeltatEstimate}
    \widehat{\Delta t}\;=\;\left(\frac{n_{+}-n_{-}}{N}\right)^{\!1/n}-t.
\end{equation}
The error in our estimate is then defined as
\begin{equation}
    \Delta t_{\mathrm{err}}=\Delta t-\widehat{\Delta t}.
\end{equation}

\begin{figure}
    \includegraphics[width=\columnwidth]{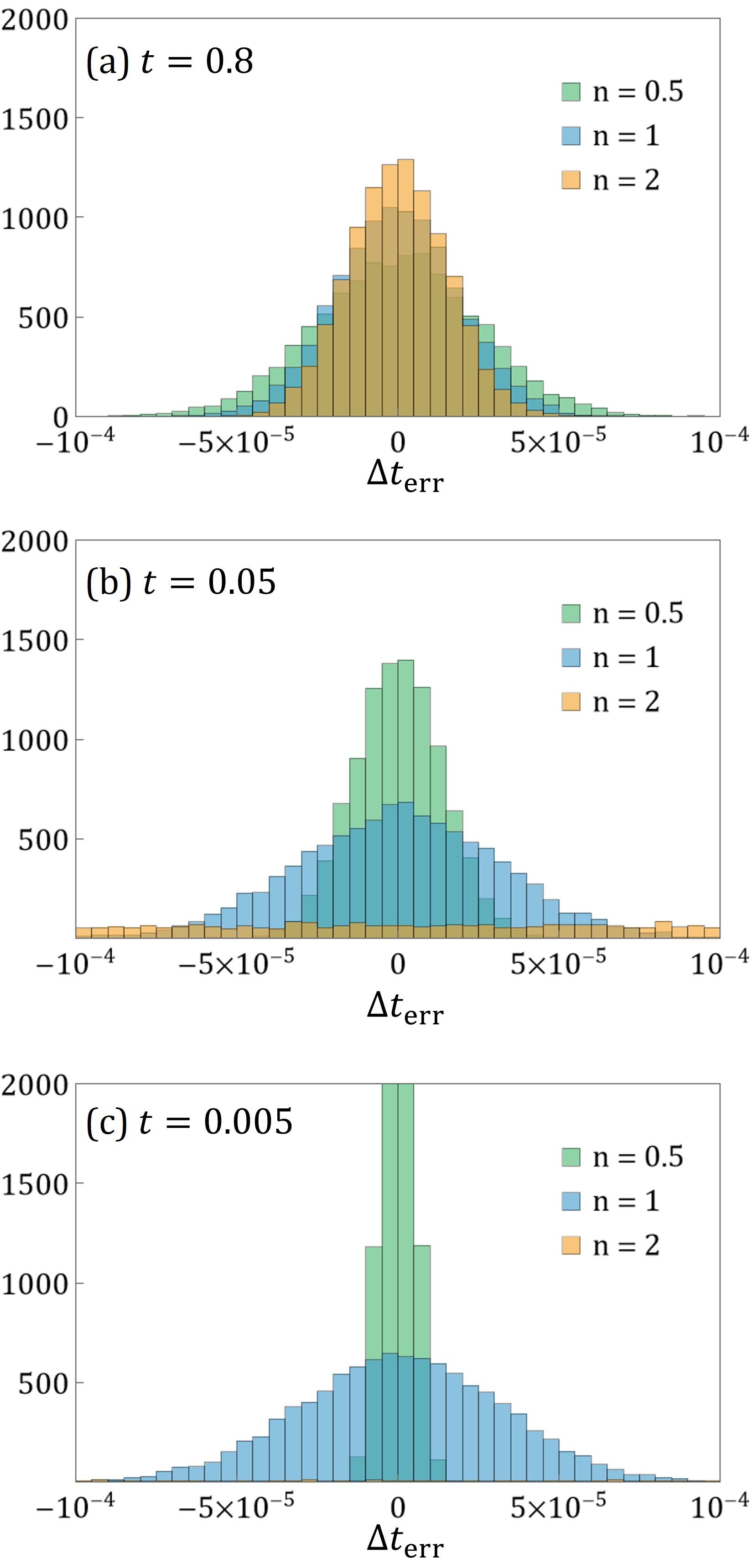}
    \caption{We simulate estimating $\Delta t$ using \cref{eq:DeltatEstimate} for $N=10^9$ measurements, and plot histograms of $\Delta t_{\mathrm{err}}$ for $10^4$ repetitions of this. We take $\Delta t=10^{-6}$. The width of the distribution should be proportional to $1/F_Q^n(t)$. In (a) we have $t=0.8$. In this case increasing $n$ increases the Fisher information, but they are all of the same order of magnitude. We can see that the histograms are Gaussian shapes, and as expected slightly narrow as $n$ increases. (b) The low transmission regime $t=0.05$. Now the QFI for $n=2$ is zero, while $n=0.5$ has higher QFI than $n=1$. As expected the histogram of errors for $n=2$ is very broad, and the narrowest distribution corresponds to $n=0.5$. (c) Very low transmission $t=0.005$. The distribution for $n=1$ is approximately the same as before, corresponding to the constant QFI. The distribution for $n=0.5$ is sharply narrowed, and truncated in this diagram. The distribution for $n=2$ is now so broad that it lies close to the horizontal axis, corresponding to its vanishing Fisher information.}\label{fig:Histogram}
\end{figure}

We simulate this for three values of $t$ in \cref{fig:Histogram}, and find results consistent with the plots of the QFI in \cref{singpass} (a). For high transmission, the error in estimating $\Delta t$ is approximately the same, though slightly narrower for $n=2$. For low transmission however we see that for $n=0.5$ the distribution of errors is significantly narrower. We also see how two-pass setups significantly reduce the information gained from samples with low transmission, compared to a single pass. 

\begin{figure*}
    \centering
    \includegraphics[width=1\linewidth]{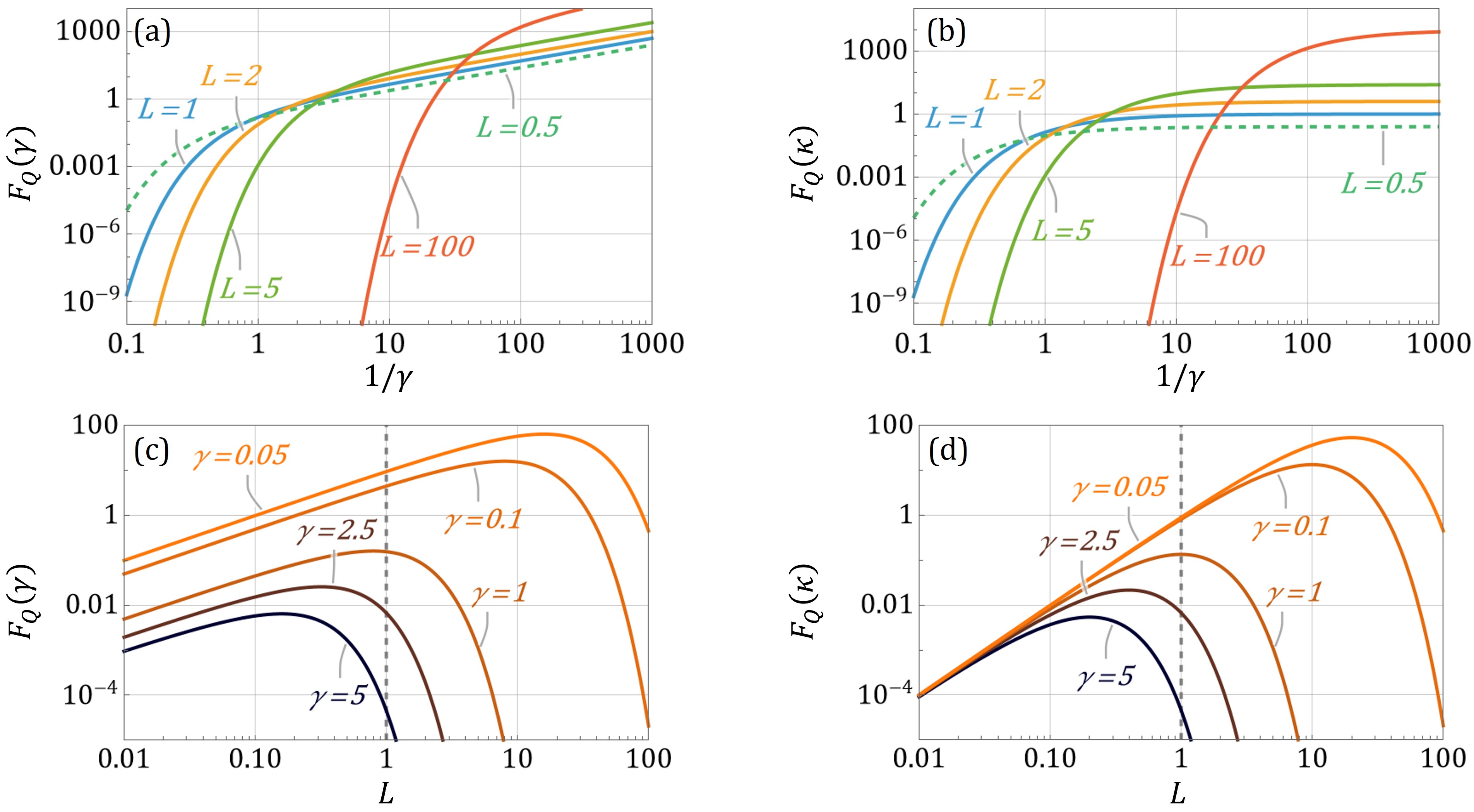}
    \caption{The QFIs for estimating material parameters $\gamma$ and $\kappa$. In (a) and (b), we plot these as functions of $1/\gamma$. QFI increases with increasing transmission (i.e. as $1/\gamma$ increases). In (c) and (d) we plot these as a function of the path length L. Increasing transmission  (decreasing $\gamma$) always increases the QFI. For each value of $\gamma$, there is a single optimal path length $L$. Unlike when parameterising in terms of $t$, now there is no change in behaviour at $L=1$.}\label{fig:MaterialParameters}
\end{figure*}

\section{Estimating material coefficients}\label{app:MaterialCoefficients}
Suppose the idler beam has effective path length $L$ through the sample. Then we can write the transmission and phase shift in terms of material parameters as
\begin{align}
    t &= e^{-\gamma L}, \\
    \phi &= \kappa L,
\end{align}
where $\gamma$ and $\kappa$ quantify the absorption and phase shift per unit length. Let us consider the QFI for estimating these parameters. Now $L$ is playing the role of our multipass parameter $n$.

In terms of these, the quantum state \cref{eq:rho_signal_3state} becomes
\begin{equation} \label{eq:rho_gk}
    \rho
    =\frac12
    \begin{pmatrix}
    1 & e^{-(\gamma+i\kappa) L}\\
    e^{-(\gamma-i\kappa) L} & 1
    \end{pmatrix}.
\end{equation}
We can write this in Bloch form,
\begin{equation}
\rho=\frac12\Bigl(\mathbb{I}+\mathbf{r}\cdot\boldsymbol{\sigma}\Bigr),
\end{equation}
where $\boldsymbol{\sigma}=(\sigma_x,\sigma_y,\sigma_z)$ is the vector of Pauli matrices, and the Bloch vector is
\begin{equation}\label{eq:bloch_vec}
    \mathbf{r}=
    \Bigl(e^{-\gamma L}\cos(\kappa L),\; e^{-\gamma L}\sin(\kappa L),\;0\Bigr),
\end{equation}
with norm $|\mathbf{r}|=e^{-\gamma L}$.

From \cref{eq:QFIBlochVector}, the QFI can be found in terms of derivatives of the Bloch vector. These are
\begin{align}
    \partial_\gamma \mathbf{r} &= -L\,\mathbf{r},\\
    \partial_\kappa \mathbf{r} &= Le^{-\gamma L}(-\sin(\kappa L),\cos(\kappa L),0).
\end{align}
This gives us
\begin{align}
    F_Q(\gamma) &=\frac{L^2}{e^{2\gamma L}-1}, \\
    F_Q(\kappa) &= L^2 e^{-2\gamma L}.
\end{align}
We plot these in \cref{fig:MaterialParameters}, as functions of both $1/\gamma$ and $L$. We see that now both absorption and phase shift show similar trends.

\clearpage
\onecolumngrid
\begin{center}
  \textbf{\large Supplementary Material for ``Ultimate sensitivity of multiparameter\\
  estimation in quantum sensing with undetected photons''}
\end{center}
\vspace{1.5em}

\setcounter{section}{0}
\setcounter{equation}{0}
\setcounter{figure}{0}
\setcounter{table}{0}
\renewcommand{\thesection}{S\arabic{section}}
\renewcommand{\thesubsection}{\thesection.\arabic{subsection}}
\renewcommand{\theequation}{S\arabic{equation}}
\renewcommand{\thefigure}{S\arabic{figure}}
\renewcommand{\thetable}{S\arabic{table}}

\section{Density Matrix Derivation}
\label{app:intensityderivation}

Here, we work in the \emph{single-pair manifold} and restrict to the \emph{four-dimensional} subspace actually populated by
the experiment/model. Concretely, the signal is restricted to the ``which-crystal'' single-photon space
$\mathcal H_s^{(1)}=\mathrm{span}\{\ket{10}_s,\ket{01}_s\}$ and the idler$+$environment sector is reduced to
the two orthogonal outcomes ``matched'' versus ``leaked'' after $n$ passes. 

We model the $n$-pass action of the object on the idler of path~$2$ via its unitary dilation:
\begin{align}
\mathcal O^{\,n}\!\left(\ket{1_{i_1}}\!\otimes\!\ket{0_E}\right)
= t^{n} e^{i n\phi}\,\ket{1_{i_1}}\!\otimes\!\ket{0_E}
+ \sqrt{1-t^{2n}}\,\ket{0^{(n)}_{i_1}}\!\otimes\!\ket{1_E},
\label{eq:npass-dilation}
\end{align}
with $\bra{1_{i_1}}\ket{0^{(n)}_{i_1}}=0$ and $\bra{0_E}\ket{1_E}=0$.

Let the post-object \emph{pure} state of \emph{signal} $\otimes$ (\emph{idler+env}) be
\begin{align}
\ket{\Psi_{\mathrm{post}}}
=\frac{1}{\sqrt2}\Big(
\ket{01}_s\!\otimes\!\ket{1_{i_1},0_E}
+ e^{i\theta}\ket{10}_s\!\otimes\!
\big[t^{n} e^{i n\phi}\,\ket{1_{i_1},0_E}
+ \sqrt{1-t^{2n}}\,\ket{0^{(n)}_{i_1},1_E}\big]
\Big),
\label{eq:psi-post}
\end{align}
where $\ket{1_{i_1},0_E}\equiv\ket{1_{i_1}}\!\otimes\!\ket{0_E}$ and
$\ket{0^{(n)}_{i_1},1_E}\equiv\ket{0^{(n)}_{i_1}}\!\otimes\!\ket{1_E}$.

Define the ordered orthonormal basis
\[
\mathcal B=\Big\{
\ket{01}\!\otimes\!\ket{1_{i_1},0_E},\;
\ket{01}\!\otimes\!\ket{0^{(n)}_{i_1},1_E},\;
\ket{10}\!\otimes\!\ket{1_{i_1},0_E},\;
\ket{10}\!\otimes\!\ket{0^{(n)}_{i_1},1_E}
\Big\}.
\]
Then
\begin{align}
\rho_{\mathrm{full}}(n)
=\ket{\Psi_{\mathrm{post}}}\!\bra{\Psi_{\mathrm{post}}}
=\frac12
\begin{pmatrix}
1 & 0 & e^{-i\theta}\,t^{n}e^{-i n\phi} & e^{-i\theta}\,\sqrt{1-t^{2n}} \\
0 & 0 & 0 & 0 \\
e^{+i\theta}\,t^{n}e^{+i n\phi} & 0 & t^{2n} & t^{n}e^{+i n\phi}\sqrt{1-t^{2n}} \\
e^{+i\theta}\,\sqrt{1-t^{2n}} & 0 & t^{n}e^{-i n\phi}\sqrt{1-t^{2n}} & 1-t^{2n}
\end{pmatrix}.
\label{eq:rho-full-matrix}
\end{align}

We now compute the reduced density matrix of the \emph{signal} by taking the partial trace
over idler+environment:
\begin{align}
\rho_s\;=\;\mathrm{Tr}_{i,e}\big[\rho_{\mathrm{full}}(n)\big].
\end{align}
Using orthogonality $\bra{1_{i_1},0_E}\ket{0^{(n)}_{i_1},1_E}=0$ and normalization,
only the diagonal idler--environment blocks contribute to the trace. In operator form,
starting from \cref{eq:psi-post},
\begin{align}
\rho_{\mathrm{full}}(n)
&=\frac12\Big(
\ket{0}\!\bra{0}\otimes\ket{1_{i_1},0_E}\!\bra{1_{i_1},0_E}
+\ket{1}\!\bra{1}\otimes\ket{\chi}\!\bra{\chi}
\nonumber\\[-2pt]
&\hspace{2.5cm}
+e^{-i\theta}\ket{0}\!\bra{1}\otimes\ket{1_{i_1},0_E}\!\bra{\chi}
+e^{+i\theta}\ket{1}\!\bra{0}\otimes\ket{\chi}\!\bra{1_{i_1},0_E}
\Big),
\end{align}
where
\(
\ket{\chi}
= t^{n} e^{i n\phi}\,\ket{1_{i_1},0_E}
+ \sqrt{1-t^{2n}}\,\ket{0^{(n)}_{i_1},1_E}.
\)

Taking the partial trace gives
\begin{align}
\rho_s
&=\frac12\Big(
\ket{0}\!\bra{0}+\ket{1}\!\bra{1}
+ e^{-i\theta}\mathrm{Tr}_{i,e}\!\big[\ket{1_{i_1},0_E}\!\bra{\chi}\big]\ket{0}\!\bra{1}
+ e^{+i\theta}\mathrm{Tr}_{i,e}\!\big[\ket{\chi}\!\bra{1_{i_1},0_E}\big]\ket{1}\!\bra{0}
\Big)\nonumber\\
&=\frac12\Big(
\ket{0}\!\bra{0}+\ket{1}\!\bra{1}
+ e^{-i\theta}\braket{\chi|1_{i_1},0_E}\,\ket{0}\!\bra{1}
+ e^{+i\theta}\braket{1_{i_1},0_E|\chi}\,\ket{1}\!\bra{0}
\Big).
\end{align}
Since
\(
\braket{1_{i_1},0_E|\chi}=t^{n}e^{i n\phi}
\)
and
\(
\braket{\chi|1_{i_1},0_E}=t^{n}e^{-i n\phi},
\)
we obtain, in the $\{\ket{0},\ket{1}\}$ basis,
\begin{align}
\rho_s
=\frac12
\begin{pmatrix}
1 & e^{-i\theta}\,t^{n}e^{-i n\phi}\\[3pt]
e^{+i\theta}\,t^{n}e^{+i n\phi} & 1
\end{pmatrix}.
\label{eq:rho-signal-general-theta}
\end{align}
For the commonly used interferometric choice $\theta=0$,
\begin{align}
\rho_s\Big|_{\theta=0}
=\frac12
\begin{pmatrix}
1 & t^{n}e^{-i n\phi}\\[3pt]
t^{n}e^{+i n\phi} & 1
\end{pmatrix}.
\label{eq:rho-signal-theta0}
\end{align}

\subsection{Purity Check}

The purity of a quantum state is defined by
\(
  \gamma=\mathrm{Tr}[\rho^{2}],
\)
with \(0<\gamma\le1\); the upper bound is reached \emph{iff} the state is
pure.  For the two–level state
\[
  \rho(t,\phi)=\frac12
  \begin{pmatrix}
    1 & t^n\,e^{-in\phi}\\
    t^n\,e^{in\phi} & 1
  \end{pmatrix},
  \qquad 0\le t\le1,
\]
purity is readily obtained:

\[
  \rho^{2}
  =\frac14
  \begin{pmatrix}
     1+t^{2n} & 2t^n\,e^{-in\phi}\\[4pt]
     2t^n\,e^{in\phi} & 1+t^{2n}
  \end{pmatrix},
  \quad
  \gamma=\mathrm{Tr}[\rho^{2}]
  =\tfrac12\bigl(1+t^{2n}\bigr).
\]

\begin{itemize}
  \item \textbf{\(t=0\) (maximal decoherence).}  
        \(\gamma=\tfrac12\); the state is completely mixed, \(
        \rho=\tfrac12(\ket{0_s}\bra{0_s}+\ket{1_s}\bra{1_s})\).
  \item \textbf{\(t=1\) (full coherence).}  
        \(\gamma=1\); the state is pure and reduces to the interference
        basis \(\tfrac12(\ket{0_s}\!\pm e^{i\phi}\ket{1_s})\).
\end{itemize}

\subsection{Measurement }
\begin{equation}
    \begin{aligned}
        \rho_s(t,\phi) &= \frac{1}{2}\left(\ket{0_s}\!\bra{0_s}+\ket{1_s}\!\bra{1_s}\right. \\
            &\phantom{=}\hspace{3em}\left.+ t^n e^{in\phi}\ket{0_s}\!\bra{1_s} + t^n e^{-in\phi}\ket{1_s}\!\bra{0_s}\right).
    \end{aligned}
\end{equation}

The interferometric projection state with a controllable phase~$\theta$
between the arms is
\[
  \ket{\psi(\theta)}
   =\tfrac1{\sqrt2}\bigl(\ket{0_s}+e^{i\theta}\ket{1_s}\bigr),
  \qquad
  \Pi(\theta)\equiv\ket{\psi(\theta)}\!\bra{\psi(\theta)}.
\]

\[
\begin{aligned}
  \Pi(\theta)
  &=\tfrac12\Bigl(
       \ket{0_s}\!\bra{0_s}
      +e^{-i\theta}\ket{0_s}\!\bra{1_s}
      +e^{i\theta}\ket{1_s}\!\bra{0_s}
      +\ket{1_s}\!\bra{1_s}
    \Bigr).
\end{aligned}
\]

Because the basis $\{\ket{0_s},\ket{1_s}\}$ is orthonormal,
$\mathrm{Tr}[\ket{s_i}\!\bra{s_j}]=\delta_{ij}$, so
only \emph{diagonal} outer products survive in
$\mathrm{Tr}\!\bigl[\rho_s\,\Pi(\theta)\bigr]$:

\[
\begin{aligned}
  I(\theta)
  &=\mathrm{Tr}\!\bigl[\rho_s(t,\phi)\,\Pi(\theta)\bigr]\\[4pt]
  &=\tfrac12\mathrm{Tr}\!\Bigl[
       \bigl(\ket{0_s}\!\bra{0_s}+\ket{1_s}\!\bra{1_s}\bigr)
       \Pi(\theta)
     \Bigr]\\
  &+\tfrac12 t^n\,
     \mathrm{Tr}\!\Bigl[
       e^{ -in\phi}\ket{0_s}\!\bra{1_s}\,\Pi(\theta)
      +e^{in\phi}\ket{1_s}\!\bra{0_s}\,\Pi(\theta)
     \Bigr].
\end{aligned}
\]

Evaluate each contribution:

\[
\begin{aligned}
  &\mathrm{Tr}\!\bigl[\ket{0_s}\!\bra{0_s}\,\Pi(\theta)\bigr]
    =\tfrac12,
  \qquad
  \mathrm{Tr}\!\bigl[\ket{1_s}\!\bra{1_s}\,\Pi(\theta)\bigr]
    =\tfrac12,\\[6pt]
  &\mathrm{Tr}\!\bigl[\ket{0_s}\!\bra{1_s}\,\Pi(\theta)\bigr]
    =\tfrac12\,e^{-i\theta},\qquad
  \mathrm{Tr}\!\bigl[\ket{1_s}\!\bra{0_s}\,\Pi(\theta)\bigr]
    =\tfrac12\,e^{ i\theta}.
\end{aligned}
\]

\[
\begin{aligned}
  I(\theta)
   &=\tfrac12\!\Bigl(\tfrac12+\tfrac12\Bigr)
    +\tfrac12 t^n\,
      \tfrac12\Bigl(e^{-in\phi}e^{-i\theta}+e^{in\phi}e^{ i\theta}\Bigr)\\[4pt]
   &=\tfrac12
    \;+\;\tfrac12\,t^n\cos(\theta+n\phi)\\[6pt]
   &\boxed{I(\theta)=\tfrac12\bigl[1+t^n\cos(\theta+n\phi)\bigr].}
\end{aligned}
\]

The constant term $1/2$ is the incoherent
background; the visibility~$t^n$ multiplies the cosine interference term
and encodes object transmissivity (or coherence amplitude) while the
phase shift~$\phi$ sets the fringe origin.

\section{QFI Calculations}
\label{qfi_calc}

We provide detailed calculations here the QFI matrix of $t$ and $\phi$. 

\[
  \bigl[\mathcal F\bigr]_{ij}
  \;=\;
  \partial_{x_i}\boldsymbol r . \partial_{x_j}\boldsymbol r
  \;+\;
  \frac{(\boldsymbol r\!\cdot\!\partial_{x_i}\boldsymbol r)\,(\boldsymbol r\!\cdot\!\partial_{x_j}\boldsymbol r)}
       {\,1-|\boldsymbol r|^{2}},
  \, x_i,x_j\in\{t,\phi\}.
\]

\[
\begin{aligned}
  \partial_t\boldsymbol r &= \bigl(n t^{\,n-1}\cos n\phi,\;n t^{\,n-1}\sin n\phi,\,0\bigr),\\[0.3em]
  \partial_\phi\boldsymbol r &= \bigl(-n t^{\,n}\sin n\phi,\;n t^{\,n}\cos n\phi,\,0\bigr);
\end{aligned}
\]
\[
\begin{aligned}
  \partial_t\boldsymbol r\!\cdot\!\partial_t\boldsymbol r &= n^{2}t^{\,2n-2},\\[0.2em]
  \partial_\phi\boldsymbol r\!\cdot\!\partial_\phi\boldsymbol r &= n^{2}t^{\,2n},\\[0.2em]
  \partial_t\boldsymbol r\!\cdot\!\partial_\phi\boldsymbol r &=0;
\end{aligned}
\qquad
\begin{aligned}
  \boldsymbol r\!\cdot\!\partial_t\boldsymbol r &= n t^{\,2n-1},\\[0.2em]
  \boldsymbol r\!\cdot\!\partial_\phi\boldsymbol r &= 0,\\[0.2em]
  1-|\boldsymbol r|^{2} &= 1-t^{\,2n}.
\end{aligned}
\]

\[
\begin{aligned}
 \mathcal F_{tt} &=
    n^{2}t^{\,2n-2}
    \;+\;
    \frac{\bigl(n t^{\,2n-1}\bigr)^{2}}{1-t^{\,2n}}
    \;=\;
    \boxed{\dfrac{n^{2}\,t^{\,2n-2}}{1-t^{\,2n}}},\\[0.8em]
 \mathcal F_{\phi\phi} &=
    n^{2}t^{\,2n}
    \;+\;0
    \;=\;
    \boxed{n^{2}\,t^{\,2n}},\\[0.8em]
 \mathcal F_{t\phi}&=\mathcal F_{\phi t}=0.
\end{aligned}
\]

\subsection{Maximizing QFI of $t$}
Let $a \equiv \ln t < 0$. Then $t^{2n} = e^{2an}$. Rewrite $f(n)$ as
\[
    f(n) = \frac{n^2 e^{2an-2a}}{1 - e^{2an}}.
\]
Differentiating with respect to $n$:
\begin{align}
    \frac{d f}{dn}
    &= \frac{(1-e^{2an})\frac{d}{dn}\big(n^2 e^{2an-2a}\big) - n^2 e^{2an-2a} \frac{d}{dn}(1-e^{2an})}{(1-e^{2an})^2} \\
    &= \frac{ e^{2an-2a} \Big[2n(1-e^{2an}) + 2n^2 a \Big] }{(1-e^{2an})^2}.
\end{align}
Setting $\frac{d f}{dn}=0$ gives
\begin{equation}
    2n(1-e^{2an}) + 2n^2 a = 0
    \;\;\Longrightarrow\;\;
    n(1 - e^{2an}) = - n^2 a.
\end{equation}
For $n>0$, we divide by $n$:
\begin{equation}
    1 - e^{2an} = -n a \quad \Longrightarrow \quad 1 + n a = e^{2an}.
\label{eq:stationary}
\end{equation}

\subsection{Maximizing QFI of $\phi$}
Fix a transmissivity \(t\in(0,1)\) and define
\[
  g(n)\;=\;n^{2}t^{2n}, 
  \qquad n>0.
\]
Write \(t^{2n}=e^{2a n}\) with \(a=\ln t<0\); then
\[
  g(n)=n^{2}e^{2a n}.
\]

\[
  \frac{\mathrm{d}g}{\mathrm{d}n}=e^{2a n}\bigl(2n+2a n^{2}\bigr)
  =2n e^{2a n}\bigl(1+a n\bigr).
\]

Besides the trivial root \(n=0\), the critical point satisfies \(1+a n=0\), giving
\[
  n_{\max}(t)= -\frac{1}{\ln t}\;>\;0
  \quad(\text{because } 0<t<1).
\]

\section{SLD Calculations}
\label{sld_calc}

\subsection{Model and Bloch representation}
We work with
\[
\rho(t,\phi)=\tfrac12
\begin{pmatrix}
1 & t^{n}e^{-in\phi}\\
t^{n}e^{+in\phi} & 1
\end{pmatrix}
=\tfrac12\bigl(I+\mathbf r\cdot\boldsymbol\sigma\bigr),
\qquad
\mathbf r=\bigl(t^{n}\cos n\phi,\;t^{n}\sin n\phi,\;0\bigr).
\]
Define the planar Pauli directions
\[
\sigma_\phi:=\cos(n\phi)\,\sigma_x+\sin(n\phi)\,\sigma_y,
\qquad
\sigma_{\phi+\frac{\pi}{2}}:=-\sin(n\phi)\,\sigma_x+\cos(n\phi)\,\sigma_y,
\]
and their unit vectors
\[
\hat{\mathbf n}_t:=\bigl(\cos n\phi,\;\sin n\phi,\;0\bigr),
\qquad
\hat{\mathbf m}_\phi:=\bigl(-\sin n\phi,\;\cos n\phi,\;0\bigr),
\]
so that $\sigma_{\hat{\mathbf n}_t}=\sigma_\phi$ and $\sigma_{\hat{\mathbf m}_\phi}=\sigma_{\phi+\frac{\pi}{2}}$.

\subsection{SLD Calculations}
For any parameter $\theta$, use the ansatz
\(
L_\theta=\alpha_\theta\,\mathbb I+\boldsymbol\beta_\theta\!\cdot\!\boldsymbol\sigma
\),
and insert it into
\(
\partial_\theta\rho=\tfrac12(\rho L_\theta+L_\theta\rho)
\).
Matching identity and vector parts gives
\begin{equation}\label{eq:system-new}
\alpha_\theta+\mathbf r\!\cdot\!\boldsymbol\beta_\theta=0,\qquad
\boldsymbol\beta_\theta+\alpha_\theta\,\mathbf r=\partial_\theta\mathbf r .
\end{equation}
Solving \cref{eq:system-new} yields the unique SLD
\begin{align}\label{eq:SLD-general-new}
\boxed{\;
\begin{aligned}
\alpha_\theta&=-\frac{\mathbf r\!\cdot\!\partial_\theta\mathbf r}{1-|\mathbf r|^2},\\[4pt]
\boldsymbol\beta_\theta&=\partial_\theta\mathbf r+
\frac{\mathbf r\!\cdot\!\partial_\theta\mathbf r}{1-|\mathbf r|^2}\,\mathbf r .
\end{aligned}}
\end{align}

With $\mathbf r=(t^{n}\cos n\phi,\;t^{n}\sin n\phi,\,0)$,
\[
\partial_t\mathbf r=n\,t^{\,n-1}(\cos n\phi,\;\sin n\phi,\,0),\qquad
\partial_\phi\mathbf r=n\,t^{\,n}(-\sin n\phi,\;\cos n\phi,\,0),
\]
and hence
\[
D_t:=\mathbf r\!\cdot\!\partial_t\mathbf r=n\,t^{\,2n-1},\qquad
D_\phi:=\mathbf r\!\cdot\!\partial_\phi\mathbf r=0,\qquad
1-|\mathbf r|^2=1-t^{2n}.
\]

\paragraph{SLD with respect to $t$.}
\[
\alpha_t=-\frac{n\,t^{2n-1}}{1-t^{2n}},\qquad
\boldsymbol\beta_t=\frac{n\,t^{\,n-1}}{1-t^{2n}}\,\hat{\mathbf n}_t.
\]
\[
\boxed{
L_t=-\frac{n\,t^{2n-1}}{1-t^{2n}}\,\mathbb I
+\frac{n\,t^{\,n-1}}{1-t^{2n}}\,\sigma_{\hat{\mathbf n}_t}}
\]

\paragraph{SLD with respect to $\phi$.}
\[
\alpha_\phi=0,\qquad
\boldsymbol\beta_\phi=n\,t^{\,n}\,\hat{\mathbf m}_\phi.
\]
\[
\boxed{
L_\phi=\;n\,t^{\,n}\,\sigma_{\hat{\mathbf m}_\phi}}
\]

Using $[\sigma_{\hat{\mathbf n}_t},\sigma_{\hat{\mathbf m}_\phi}]=2i\,\sigma_z$,
\begin{align}
[L_t,L_\phi]
&= \frac{n\,t^{\,n-1}}{1-t^{2n}}\; n\,t^{\,n}\,
   [\sigma_{\hat{\mathbf n}_t},\sigma_{\hat{\mathbf m}_\phi}]\\
&= \frac{n^{2}\,t^{\,2n-1}}{1-t^{2n}}\,
   [\sigma_{\hat{\mathbf n}_t},\sigma_{\hat{\mathbf m}_\phi}]\\
&= \frac{n^{2}\,t^{\,2n-1}}{1-t^{2n}}\,(2i\,\sigma_z)\\
&= \boxed{\frac{2i\,n^{2}\,t^{\,2n-1}}{1-t^{\,2n}}\,\sigma_z}\,.
\end{align}

\subsection{Eigen\-decomposition of the SLD operators}
\label{SLD_vec}

\subsubsection{Saturating QFI of $t$ from Eigenvectors and eigenvalues of $L_t$}
Write $L_t=a_t\mathbb I+b_t\,\sigma_{\hat{\mathbf n}_t}$ with
\[
a_t=-\frac{n\,t^{2n-1}}{1-t^{2n}},\qquad
b_t=\frac{n\,t^{\,n-1}(1+t^{2n})}{1-t^{2n}}\ (>0).
\]
The eigenvalues are
\[
\boxed{\;
\lambda_{t,\pm}=a_t\pm b_t\; }.
\]
The eigenvectors of $\sigma_{\hat{\mathbf n}_t}$ are
\[
\boxed{\;
\ket{t_+}
=\frac{1}{\sqrt2}\!\begin{pmatrix}1\\ e^{+i n\phi}\end{pmatrix},
\qquad
\ket{t_-}
=\frac{1}{\sqrt2}\!\begin{pmatrix}1\\ -\,e^{+i n\phi}\end{pmatrix}}
\]
and satisfy
\(
L_t\ket{t_{\pm}}=\lambda_{t,\pm}\ket{t_{\pm}}.
\)

The eigenvectors of $L_t$ are 
\(
\{\ket{t_+},\ket{t_-}\}
\)
with projectors  
\(
\Pi_\pm=\ket{t_{\pm}}\!\bra{t_{\pm}}.
\)

The outcome probabilities are given by- 
\[
p_\pm(t)=\mathrm{Tr}[\rho(t,\phi)\,\Pi_\pm]
        =\frac{1\pm r}{2},\quad
\partial_t p_\pm=\pm\frac{\dot r}{2}.
\]

For a two-outcome projective measurement we have the classical fisher information $F_C^{(t)}$ given by
\begin{align}
F_C^{(t)}=\sum_{\pm}\frac{(\partial_t p_\pm)^2}{p_\pm}
          =\frac{\dot r^{\,2}}{4}\!
            \left(\frac{1}{p_+}+\frac{1}{p_-}\right)\nonumber \\
          =\frac{\dot r^{\,2}}4\!
            \left(\frac{2}{1+r}+\frac{2}{1-r}\right)
          =\frac{\dot r^{\,2}}{1-r^{2}} .
\end{align}

Inserting $\dot r=n\,t^{\,n-1}$ and $r^2=t^{2n}$ we have
\[
\boxed{\;
F_C^{(t)}=\frac{n^{2}t^{2n-2}}{1-t^{2n}}
        =\mathcal F_t}\; .
\]

Thus the SLD basis for $t$ indeed achieves the quantum limit.

\subsubsection{Saturating QFI of $\phi$ from eigenvectors and eigenvalues of $L_\phi$}
Here $L_\phi=\gamma\,\sigma_{\hat{\mathbf m}_\phi}$ with $\gamma=n\,t^n$.
Thus
\[
\boxed{\;\lambda_{\phi,\pm}=\pm\,\gamma=\pm\,n\,t^{\,n}\;}
\]
and the eigenvectors of $\sigma_{\hat{\mathbf m}_\phi}$ are

\begin{equation}\boxed{
\ket{\phi_-}
=\frac{1}{\sqrt2}\!\begin{pmatrix}1\\[2pt] -i\,e^{i n \phi} \end{pmatrix},
\quad
\ket{\phi_+}
=\frac{1}{\sqrt2}\!\begin{pmatrix}1\\ i\,e^{i n\phi}\end{pmatrix}}
\end{equation}

so that
\(
L_\phi\ket{\phi_{\pm}}=\lambda_{\phi,\pm}\ket{\phi_{\pm}}.
\) which satisfy
\begin{equation}
L_\phi\,\ket{\phi_{\pm}}=\lambda_{\phi,\pm}\,\ket{\phi_{\pm}}.
\end{equation}

Let $\mathbf r(\phi)$ denote the Bloch vector of $\rho(\phi)$:
\begin{equation}
\mathbf r(\phi)=\big(t^{n}\cos(n\phi),\, t^{n}\sin(n\phi),\,0\big).
\end{equation}
For any qubit projective POVM $\Pi_\pm=\tfrac12\!\left(\mathbb I\pm \hat{\mathbf m}\!\cdot\!\boldsymbol\sigma\right)$, the Born rule gives
\begin{equation}
p_\pm(\phi)=\mathrm{Tr}\!\big[\rho(\phi)\Pi_\pm\big]
=\tfrac12\!\left(1\pm \hat{\mathbf m}\!\cdot\!\mathbf r(\phi)\right).
\end{equation}

To attain the SLD quantum limit locally, one fixes the two–outcome POVM to the SLD–optimal measurement \emph{at the operating point} $\phi_0$, i.e.\ use $\{\Pi_{\pm}^{(\phi_0)}\}$ determined by the SLD eigenstates evaluated at $\phi_0$. With this \emph{fixed} POVM, the outcome probabilities as $\phi = \phi_0 +  \delta \phi$ varies are
\begin{equation}
\boxed{\,p_\pm(\phi)=\tfrac12 \pm \tfrac12\,t^{n}\,\sin\!\big(n \,\delta \phi\big)\, }.
\end{equation}
In particular, at the operating point,
\begin{equation}
\boxed{\,p_\pm(\phi_0)=\tfrac12\, }.
\end{equation}

Differentiating the fixed–POVM probabilities with respect to $\phi$,
\begin{equation}
\partial_\phi p_\pm(\phi) =\pm \frac{n}{2}\,t^{n}\,\cos\!\big(n \,\delta \phi\big),
\end{equation}
\begin{equation}
\Rightarrow \boxed{\,\partial_\phi p_\pm\big|_{\phi_0}=\pm \tfrac{n}{2}\,t^{n}\, }.
\end{equation}

The (scalar) classical Fisher information (CFI) of this two–outcome measurement, evaluated at $\phi_0$, is
\begin{align}
F_C^{\phi}\Big|_{\phi_0}
&=\sum_{\pm}\frac{\big(\partial_\phi p_\pm\big)^2}{p_\pm}\Bigg|_{\phi_0}
=2\cdot\frac{\left(\tfrac{n}{2}t^{n}\right)^{\!2}}{\tfrac12}
=\boxed{\,n^{2}\,t^{2n}\,}.
\end{align}
This equals the SLD quantum Fisher information (QFI) for the family $\rho(\phi)$, confirming that the SLD–derived projective POVM fixed at $\phi_0$ is \emph{locally} optimal and saturates the quantum Cramér–Rao bound at that operating point.

\section{Calculation of Holevo Bound and Nagaoka Bound}\label{app:HolevoNagaoka}

We have the density Matrix given by 
\begin{align}
\rho
&= \frac12
\begin{pmatrix}
1 & t^{n} e^{-i n\phi}\\
t^{n} e^{i n\phi} & 1
\end{pmatrix}.
\end{align}

Now we assume we have fixed $t$ and $\phi$. Our goal is not to make small changes to it. Let's assume that these small changes are $\tau$ and $\theta$. We 

\begin{align}
\rho = 
\begin{pmatrix}
 \frac{1}{2} & \frac{1}{2} e^{-i n (\theta +\phi )} (t+\tau )^n \\
 \frac{1}{2} e^{i n (\theta +\phi )} (t+\tau )^n & \frac{1}{2} \\
\end{pmatrix}
\end{align}

We also know from above calculations the derivatives becode
\begin{align}
 \partial_{\tau}\rho =    \begin{pmatrix}
0 & \dfrac{n}{2}\, e^{-i n (\theta+\phi)} (t+\tau)^{\,n-1} \\
\dfrac{n}{2}\, e^{i n (\theta+\phi)} (t+\tau)^{\,n-1} & 0
\end{pmatrix} 
\end{align}

\begin{align}
    \partial_{\phi}\rho = \begin{pmatrix}
 0 & \frac{1}{2} i n e^{-i n (\theta +\phi )} (t+\tau )^n \\
 -\frac{1}{2} i n e^{i n (\theta +\phi )} (t+\tau )^n & 0 \\
\end{pmatrix}
\end{align}

In order to calculate the Holevo Bound we define the hermitian matrices $X_t$ and $X_{\phi}$ such that the below equations hold true.

\begin{align}
    \mathrm{tr}(X_t \rho) =  \mathrm{tr}(X_{\phi} \rho) = 0 \nonumber\\
    \mathrm{tr}(X_i \partial_{j}\rho) = \delta_{ij}
    \label{xteq}
\end{align}
 where $\{i, j\} \in \{t, \phi \}$. We will consider the following
\begin{align}
    X_t = \begin{pmatrix}
        a & b+ic\\
        b-ic & d
    \end{pmatrix}\\
    X_{\phi} = \begin{pmatrix}
        e & f+ig\\
        f-ig & h
    \end{pmatrix}
\end{align}

So we can see that the we have 8 variables and the equations in\,\ref{xteq} we have 6 equations. So the above matrix can be simplified to only 2 variables. Evaluating the parameters of $X_t$ and $X_{\phi}$ we have 
 \begin{align}
     X_t = \begin{pmatrix}
 -d-\frac{2 (t+\tau )}{n} & \frac{e^{-i n (\theta +\phi )} (t+\tau )^{1-n}}{n} \\
 \frac{e^{i n (\theta +\phi )} (t+\tau )^{1-n}}{n} & d \\
\end{pmatrix} \nonumber\\
X_{\phi} = \begin{pmatrix}
 -h & \frac{i e^{-i n (\theta +\phi )} (t+\tau )^{-n}}{n} \\
 -\frac{i e^{i n (\theta +\phi )} (t+\tau )^{-n}}{n} & h \\
\end{pmatrix}
\end{align}

where $\{h, d\} \in \mathbb{R}$. 

Then we define the Matrix $\mathcal{Z}$ as $\mathcal{Z}_{ij} = \mathrm{tr}(\rho X_i X_j)$

\begin{align}
    \mathcal{Z} = \begin{pmatrix}
        d^2+\frac{2 d (t+\tau )}{n}+\frac{(t+\tau )^{2-2 n}}{n^2} & \frac{(h n-i) (d n+t+\tau )}{n^2}\\
        \frac{(h n+i) (d n+t+\tau )}{n^2} & h^2+\frac{(t+\tau )^{-2 n}}{n^2}
    \end{pmatrix}
\end{align}

\subsection{Holevo Bound}
\label{hol_app}
So now we have the HCRB as given we have the weight matrix $W = \mathbb{I}$ - 
\begin{equation}
    \mathcal{C}_H = \min_{X} \mathrm{Tr}[\mathcal{Z}[X]] + \mathrm{Tr}(\mathrm{Abs}[\mathrm{Im}(\mathcal{Z}[X])])
\end{equation}

\begin{align}
    H = \mathrm{tr}(\mathcal{Z}) + \mathrm{Tr}(\mathrm{Abs}[\mathrm{Im}(\mathcal{Z}[X])]) \nonumber \\
    = d^2+\frac{2 d (t+\tau )}{n}+\frac{(t+\tau )^{2-2 n}}{n^2} + h^2+\frac{(t+\tau )^{-2 n}}{n^2} + 2\left|\frac{dn+t+\tau}{n^2}\right| )  \nonumber \\
    \lim_{\tau \rightarrow 0}\, H = d^2+\frac{2 d t }{n}+\frac{(t )^{2-2 n}}{n^2} + h^2+\frac{(t )^{-2 n}}{n^2} +2\left|\frac{dn+t}{n^2}\right| )
\end{align}

So the minimum value of the Holevo bound will occur at $d = -\frac{t}{n}$ and $h = 0$ and is given by

\begin{align}
    \min_{d,h} \lim_{\tau \rightarrow 0}H\nonumber \\
    \mathcal{C}_H = \frac{1-t^{2n}}{n^2t^{2n-2}} + \frac{1}{n^2t^{2n}}
\end{align}

\subsection{Nagaoka (NCRB) for the model}
\label{nag_app}
We consider the two-parameter qubit model
\[
\rho=\tfrac12
\begin{pmatrix}
1 & t^{n}e^{-in\phi}\\
t^{n}e^{in\phi} & 1
\end{pmatrix},\qquad 0<t<1,\ \ n>0,
\]

For two parameters, the Nagaoka bound reads
\[
\mathcal{C}_N
=\min_{X}\ \underbrace{\mathrm{Tr}[\mathcal{Z}[X]]}_{\mathrm{Tr}(\rho X_t^2)+\mathrm{Tr}(\rho X_\phi^2)}
\;+\;
\underbrace{\Big\|\ \rho^{1/2}\,[X_t,X_\phi]\ \rho^{1/2}\ \Big\|_{1}}_{\text{incompatibility penalty}}.
\]

We did the detailed calculation using Mathematica\,\cite{onlinecode} and found that the minimum value of the second term occurs at $d = -t/n$ and $h=0$ (same parameters that minimizes the first term) and it gives
\begin{equation}
\boxed{\;
\bigl\|\rho^{1/2}[X_t,X_\phi]\rho^{1/2}\bigr\|_1
=\frac{2\,t^{\,1-2n}}{n^{2}}\;\sqrt{\,1-t^{2n}\,}\,,
\qquad 0<t<1,\ \ n>0.\;}
\end{equation}

The result depends only on $t$ and $n$; the phase $\phi$ drops out by rotational
symmetry in the $xy$-plane and the final expression of Nagaoka bound is given by-
\[
\mathcal{C}_N
=\frac{1+t^{2}-t^{\,2n+2}}{n^{2}\,t^{\,2n}}
+\frac{2\,t\,\sqrt{\,1-t^{\,2n}\,}}{n^{2}\,t^{\,2n}}
=\boxed{\ \frac{\big(1+t\sqrt{\,1-t^{\,2n}\,}\big)^{2}}{n^{2}\,t^{\,2n}}\ }.
\]

\subsection{Calculation of Variance}
\label{varcalc}

We also have the final values of$X_t$ and $X_{\phi}$ as
\begin{align}
    X_t = \left(
\begin{array}{cc}
 -\frac{t}{n} & \frac{t^{1-n} e^{-i n \phi }}{n} \\
 \frac{t^{1-n} e^{i n \phi }}{n} & . -\frac{t}{n} \\
\end{array}
\right)\nonumber \\
X_{\phi} = \left(
\begin{array}{cc}
 0 & \frac{i t^{-n} e^{-i n \phi }}{n} \\
 -\frac{i t^{-n} e^{i n \phi }}{n} & 0 \\
\end{array}
\right)
\end{align}

We get the Eigen Vectors of them given by
\begin{align}
    X_{t_1} = \begin{pmatrix}
        e^{-i n \phi }\\
        1
    \end{pmatrix} \qquad
    X_{t_2} = \begin{pmatrix}
        -e^{-i n \phi }\\
        1
    \end{pmatrix}\nonumber\\
    X_{\phi_1} = \begin{pmatrix}
        ie^{-i n \phi }\\
        1
    \end{pmatrix}\qquad
    X_{\phi_2} = \begin{pmatrix}
        -ie^{-i n \phi }\\
        1
    \end{pmatrix}\qquad
\end{align}

Now we have Measurement operators given by\,(after making sure $\Pi(X_{t_1}) + \Pi(X_{t_2}) + \Pi(X_{\phi_1})+ \Pi(X_{\phi_2}) = \mathcal{I}$)
\begin{align}
    \Pi(X_{t_1}) = \lambda\left(
\begin{array}{cc}
 \frac{1}{2} & \frac{1}{2} e^{-i n \phi } \\
 \frac{1}{2} e^{i n \phi } & \frac{1}{2} \\
\end{array}
\right) \qquad \Pi(X_{t_2}) = \lambda\left(
\begin{array}{cc}
 \frac{1}{2} & -\frac{1}{2} e^{-i n \phi } \\
 -\frac{1}{2} e^{i n \phi } & \frac{1}{2} \\
\end{array}
\right) \nonumber \\
\Pi(X_{\phi_1}) = (1- \lambda)\left(
\begin{array}{cc}
 \frac{1}{2} & -\frac{1}{2} i e^{-i n \phi } \\
 \frac{1}{2} i e^{i n \phi } & \frac{1}{2} \\
\end{array}
\right) \qquad \Pi(X_{\phi_2}) = (1- \lambda)\left(
\begin{array}{cc}
 \frac{1}{2} & \frac{1}{2} i e^{-i n \phi } \\
 -\frac{1}{2} i e^{i n \phi } & \frac{1}{2} \\
\end{array}
\right)
\end{align}

Now we consider the two equations below
\begin{align}
    l\, \Pi(X_{t_1}) + m \, \Pi(X_{t_2}) = \begin{pmatrix}
 \frac{\lambda  l}{2}+\frac{\lambda  m}{2} & \frac{1}{2} \lambda  l e^{-i n \phi }-\frac{1}{2} \lambda  m e^{-i n \phi } \\
 \frac{1}{2} \lambda  l e^{i n \phi }-\frac{1}{2} \lambda  m e^{i n \phi } & \frac{\lambda  l}{2}+\frac{\lambda  m}{2} \\
\end{pmatrix}
 \nonumber \\
= \begin{pmatrix}
 -\frac{t}{n} & \frac{t^{1-n} e^{-i n \phi }}{n} \\
 \frac{t^{1-n} e^{i n \phi }}{n} & -\frac{t}{n} \\
\end{pmatrix}\\
p\, \Pi(X_{\phi_1}) + q \, \Pi(X_{\phi_2}) =\begin{pmatrix}
 \frac{1}{2} (1-\lambda ) p+\frac{1}{2} (1-\lambda ) q & \frac{1}{2} i (1-\lambda ) q e^{-i n \phi }-\frac{1}{2} i (1-\lambda ) p e^{-i n \phi } \\
 \frac{1}{2} i (1-\lambda ) p e^{i n \phi }-\frac{1}{2} i (1-\lambda ) q e^{i n \phi } & \frac{1}{2} (1-\lambda ) p+\frac{1}{2} (1-\lambda ) q \\
\end{pmatrix} \nonumber \\
= \begin{pmatrix}
 0 & \frac{i t^{-n} e^{-i n \phi }}{n} \\
 -\frac{i t^{-n} e^{i n \phi }}{n} & 0 \\
\end{pmatrix}
\end{align}

Solving the above we have
\begin{align}
    l = -\frac{t^{1-n} \left(t^n+1\right)}{\lambda  n} \nonumber \\
    m = \frac{t^{1-n} \left(t^n+1\right)}{\lambda  n}-\frac{2 t}{\lambda  n} \nonumber \\
    p = \frac{t^{-n}}{(\lambda -1) n} \nonumber \\
    q = -\frac{t^{-n}}{(\lambda -1) n}
\end{align}

Now to get the variance we do
\begin{align}
    \mathrm{Tr}((l^2 \Pi(X_{t_1}) + m^2 \Pi(X_{t_2}) + p^2 \Pi(X_{\phi_1}) + q^2 \Pi(X_{\phi_2})) \rho) = \frac{t^{-2 n} \left((\lambda -1) \left(-t^2\right) \left(t^{2 n}-1\right)-\lambda \right)}{(\lambda -1) \lambda  n^2}
\end{align}

\subsection{Exact comparison of \texorpdfstring{$\mathrm{Var}(\lambda,t,n)$}{Var} and \texorpdfstring{$\mathcal{C}_N(t,n)$}{CN}}
\label{subsec:exact_Var_vs_CN}

We work under the domain
\[
0<t<1,\qquad 0<\lambda<1,\qquad n>0.
\]

\begin{align}
\mathrm{Var}(\lambda,t,n)
&=\frac{t^{-2 n}\!\left((\lambda -1)(-t^2)(t^{2 n}-1)-\lambda \right)}{(\lambda -1)\lambda\, n^2},\\[3pt]
\mathcal{C}_N(t,n)
&=\frac{\bigl(1+t\sqrt{1-t^{2n}}\bigr)^{2}}{n^{2}\,t^{2n}}.
\end{align}

Cancelling the common factor $n^{-2}t^{-2n}$, set
\[
B:=\bigl(1+t\sqrt{1-t^{2n}}\bigr)^2,\qquad D:=t^2(1-t^{2n}),
\]
and write
\begin{equation}
\label{eq:f-lambda}
f(\lambda):=\mathrm{Var}-\mathcal{C}_N
=\frac{\lambda(D-1)-D}{\lambda(\lambda-1)}-B
=\frac{-\Big(B\lambda^2-(B{+}D{-}1)\lambda+D\Big)}{\lambda(\lambda-1)}.
\end{equation}
The quadratic in the numerator has discriminant
\[
\Delta=(B{+}D{-}1)^2-4BD=0,
\]
hence it is a perfect square:
\[
B\lambda^2-(B{+}D{-}1)\lambda+D
=B\bigl(\lambda-\lambda^\star\bigr)^2,
\qquad
\lambda^\star=\frac{B+D-1}{2B}
=\frac{t\,\sqrt{1-t^{2n}}}{1+t\,\sqrt{1-t^{2n}}}\in(0,1).
\]
Since $0<\lambda<1$ implies $\lambda(\lambda-1)<0$, equation \cref{eq:f-lambda} gives
\[
f(\lambda)
=\frac{B(\lambda-\lambda^\star)^2}{\lambda(1-\lambda)}
\;\ge\;0,
\]
with equality iff $\lambda=\lambda^\star$.

\begin{equation}
\boxed{\;
\mathrm{Var}(\lambda,t,n)-\mathcal{C}_N(t,n)\ \ge\ 0
\quad\text{for }0<t<1,\ n>0,\ 0<\lambda<1,
\quad\text{and }=\ 0\iff \lambda=\frac{t\sqrt{1-t^{2n}}}{1+t\sqrt{1-t^{2n}}}.
\;}\label{optimlam}
\end{equation}

\subsection{Beating The Nagaoka Bound}
\label{nagaokabeat}
For a fixed number of passes $n>0$, we consider the single-parameter Cram\'er--Rao
costs for estimating $\phi$ and $t$,
\begin{align}
F_{\phi}(n,t) &= \frac{1}{n^{2}\,t^{2n}},\\[3pt]
F_{t}(n,t) &= \frac{1 - t^{2n}}{n^{2}\,t^{\,2n-2}},
\end{align}
and the joint (two-parameter) cost
\begin{equation}
C_{N}(n,t) \;=\; \frac{\bigl(1 + t\sqrt{1-t^{2n}}\bigr)^{2}}{n^{2}\,t^{2n}}.
\end{equation}
Given a total of $N_{\mathrm{tot}}$ independent trials, the joint protocol uses the
same measurement in every trial and yields the total variance
\begin{equation}
V_{\mathrm{joint}}(n,t) \;=\; \frac{C_{N}(n,t)}{N_{\mathrm{tot}}}.
\end{equation}

In a mixture strategy, we allocate a fraction $x\in(0,1)$ of trials to a protocol
optimized for $t$ and the remaining fraction $1-x$ to a protocol optimized for $\phi$.
Allowing distinct pass numbers $n_t$ and $n_\phi$ for these runs, the total (sum)
variance in the continuous allocation limit is
\begin{equation}
V_{\mathrm{mix}}(n_t,n_\phi,t;x)
\;=\;\frac{1}{N_{\mathrm{tot}}}
\left(\frac{F_t(n_t,t)}{x}+\frac{F_\phi(n_\phi,t)}{1-x}\right).
\end{equation}
Minimizing over $x$ is elementary\,(for now we will not consider integer regimes). For any $a,b>0$ one has
\begin{equation}
\min_{0<x<1}\left(\frac{a}{x}+\frac{b}{1-x}\right)
=\bigl(\sqrt{a}+\sqrt{b}\bigr)^{2},
\label{eq:split_identity}
\end{equation}
attained at $x^\star=\sqrt{a}/(\sqrt{a}+\sqrt{b})$. Applying \cref{eq:split_identity} with
$a=F_t(n_t,t)$ and $b=F_\phi(n_\phi,t)$ gives the $N\to\infty$ mixture cost
\begin{equation}
V_{\mathrm{mix},\infty}(t)
\;=\;\min_{n_t>0,\;n_\phi>0}\;
\frac{\Bigl(\sqrt{F_t(n_t,t)}+\sqrt{F_\phi(n_\phi,t)}\Bigr)^2}{N_{\mathrm{tot}}}.
\label{eq:Vmix_infty_def}
\end{equation}

If the mixture protocol is restricted to use the same number of passes for both
single-parameter runs, i.e.\ $n_t=n_\phi=n$, then the closed form in
\cref{eq:Vmix_infty_def} reproduces the joint cost exactly. This is what we observed in Figure 5(a). Indeed,
\begin{align}
\sqrt{F_\phi(n,t)} &= \frac{1}{n\,t^{n}},\\[3pt]
\sqrt{F_t(n,t)} &= \sqrt{\frac{1-t^{2n}}{n^2 t^{2n-2}}}
= \frac{t\sqrt{1-t^{2n}}}{n\,t^{n}},
\end{align}

\begin{align}
\Bigl(\sqrt{F_t(n,t)}+\sqrt{F_\phi(n,t)}\Bigr)^2\nonumber \\
= \frac{\bigl(1+t\sqrt{1-t^{2n}}\bigr)^2}{n^2 t^{2n}}
= C_N(n,t).
\label{eq:CN_as_mix}
\end{align}
Hence, for every $n>0$,
\begin{equation}
\min_{0<x<1} V_{\mathrm{mix}}(n,n,t;x)
=\frac{C_N(n,t)}{N_{\mathrm{tot}}}
=V_{\mathrm{joint}}(n,t).
\label{eq:diag_equals_joint}
\end{equation}

The optimized joint bound is
\begin{equation}
V_{\mathrm{joint},\infty}(t)
=\min_{n>0}\frac{C_N(n,t)}{N_{\mathrm{tot}}}.
\label{eq:Vjoint_infty_def}
\end{equation}
On the other hand, the mixture optimization in \cref{eq:Vmix_infty_def} ranges over
all $(n_t,n_\phi)$ and in particular includes the diagonal choices $(n_t,n_\phi)=(n,n)$\,Figure 5(b).
Therefore,
\begin{align}
V_{\mathrm{mix},\infty}(t)
&=\min_{n_t,n_\phi>0}\frac{\bigl(\sqrt{F_t(n_t,t)}+\sqrt{F_\phi(n_\phi,t)}\bigr)^2}{N_{\mathrm{tot}}}
\notag\\
&\le \min_{n>0}\frac{\bigl(\sqrt{F_t(n,t)}+\sqrt{F_\phi(n,t)}\bigr)^2}{N_{\mathrm{tot}}}\nonumber \\
&\overset{\cref{eq:CN_as_mix}}{=}
\min_{n>0}\frac{C_N(n,t)}{N_{\mathrm{tot}}}
=V_{\mathrm{joint},\infty}(t),
\end{align}

\end{document}